\documentclass[pra, aps, twocolumn, 
showpacs]{revtex4-1}

\usepackage[colorlinks,urlcolor=blue,citecolor=blue,linkcolor=blue]{hyperref}

\usepackage{physics}
\usepackage[english]{babel} 
\usepackage[english]{layout}
\usepackage{amsmath,amsfonts,amssymb}
\usepackage{graphicx}
\usepackage{float}
\usepackage{color}
\usepackage{braket}
\usepackage{version}
\usepackage{array,multirow,makecell}
\usepackage{afterpage}
\usepackage[toc,page]{appendix} 
\usepackage{bbold}

\newcommand{\up}{\uparrow}
\newcommand{\dn}{\downarrow}

\newcommand{\new}[1]{{\color{black}#1}}

\begin{document}
\title{Interplay between polarization and quantum correlations of confined polaritons
}

\author{O. Bleu}
\affiliation{School of Physics and Astronomy, Monash University, Victoria 3800, Australia}
\affiliation{ARC Centre of Excellence in Future Low-Energy Electronics Technologies, Monash University, Victoria 3800, Australia}

\author{J. Levinsen}
\affiliation{School of Physics and Astronomy, Monash University, Victoria 3800, Australia}
\affiliation{ARC Centre of Excellence in Future Low-Energy Electronics Technologies, Monash University, Victoria 3800, Australia}

\author{M. M. Parish}
\affiliation{School of Physics and Astronomy, Monash University, Victoria 3800, Australia}
\affiliation{ARC Centre of Excellence in Future Low-Energy Electronics Technologies, Monash University, Victoria 3800, Australia}

\begin{abstract}
\new{We investigate polariton quantum correlations in a coherently driven box cavity in the low driving regime, with a particular focus on accounting for the polarization degree of freedom.
The possibility of having different interaction strengths between co- and cross-circularly polarized polaritons as well as a realistic linear-polarization splitting allows one to model the system as two coupled nonlinear resonators with both self- and cross-Kerr-like nonlinearities, thus making our results potentially relevant for other experimental platforms.}
Within an effective wave-function approach, we obtain analytical expressions for the steady-state polarization-resolved polariton populations and second-order correlation functions, which agree very well with 
our numerical results obtained from a Lindblad master equation.
Notably, we highlight that depending on the excitation polarization (circular or linear), both the unconventional (interference-mediated) and conventional (mediated by nonlinearities) antibunchings can be investigated in a single cavity.
Moreover, using our results, 
we argue that recent experiments on confined fiber-cavity polaritons are likely to have probed a regime where the dominant interaction is between cross-polarized polaritons,  
which is characteristic of the polariton Feshbach resonance. 
We furthermore investigate the regime close to resonance using a two-channel model, and 
we show that systems with large biexciton binding energies, such as atomically thin semiconductors, are promising platforms for realizing strong polariton antibunching. 

\end{abstract}

\maketitle

\section{Introduction}

The pursuit of nonlinear optics at the few photon level has been a major goal of quantum optics in the past decades \cite{Chang2014}, and it is important for the prospect of quantum simulation \cite{Hartmann2016,Noh2016} and quantum information processing \cite{OBrien2009} with light.
A key step in this direction is the achievement of the photon blockade (PB) regime, where the presence of a single quantum of radiation in an optical resonator forbids the entrance of another. 
The PB was originally proposed for a coherently driven single-mode cavity embedding a medium with strong Kerr nonlinearities \cite{Leonnski1994,Imamoglu1997}, a regime that is sometimes called the Kerr-blockade.  
A similar PB 
can occur in two-level systems described by the Jaynes-Cummings model in the strong-coupling regime \cite{Tian1992}, and this has been realized in a range of systems such as single-atom \cite{Birnbaum2005},  semiconductor quantum dot \cite{Faraon2008}, or microwave \cite{Lang2011} cavities. PB has also been demonstrated in a cavity filled with an atomic Rydberg ensemble, which effectively corresponds to a system of three-level atoms~\cite{Jia2018}. 

In practice, the hallmark of the PB is strong antibunching in the zero-delay intensity correlation function of the emitted light, which is a characteristic of nonclassicality~\cite{Paul1982}.
While strictly speaking this requires large effective photon-photon interactions to overcome cavity losses, 
the use of coupled cavities can give rise to a blockade of a two-photon state even in weakly nonlinear systems thanks to finely tuned quantum interferences  
\cite{Liew2010,bamba2011origin,Lemonde2014,Flayac2017}. Experimental observations of this so-called unconventional blockade effect have been reported in the optical \cite{Snijders2018} and microwave domains \cite{Vaneph2018}.
Further extensions have also been proposed, such as non-reciprocal or dynamical blockade effects for single-mode Kerr resonators \cite{Huang2018,Ghosh2019}, or the use of ``phase space filling saturation'' in semiconductors to provide an effective nonlinearity \cite{Kyriienko2020}. 
\new{Photon antibunching in a given optical mode has also been interpreted in terms of squeezed Gaussian states for several weakly driven systems~\cite{Lemonde2014,Zubizarreta2020a,Zubizarreta2020b}.}

A promising route to achieving the conventional photon blockade is the use of strongly coupled exciton-polaritons in a semiconductor pillar or box cavity~\cite{Verger2006}.
In the context of inorganic semiconductor cavities, the relevant nonlinearity is given by the polariton-polariton interaction strength. 
Recently, signatures towards the realization of the PB
have been reported in intensity correlation measurements in high-finesse fiber cavities \cite{delteil2019,MunozMatutano2019,Gerace2019}.
These studies, as well as the original proposal \cite{Verger2006}, modelled the cavity as a single mode optical resonator that neglected the polarization degree of freedom.

However, polarization is known to play an important role in semiconductor microcavities \cite{shelykh2009polariton}.
In particular, polaritons carry a pseudospin associated with the photon circular polarization and the quantum-well exciton spin, which gives rise to two different interaction strengths for co- and cross-circularly polarized polaritons 
\cite{Bleu2020}.
At the mean-field level, this characteristic has led to the prediction and observation of polarization multistability for resonantly pumped condensates \cite{Gippius2007,Shelykh2008,Adrados2010,Amo2010,Paraiso2010,Sarkar2010} and it is also relevant for condensates formed under nonresonant pump \cite{Laussy2006,Kasprzak2007,shelykh2009polariton,Ohadi2015}.
While the inter-spin interaction strength is typically assumed to be attractive and smaller than the triplet one, in principle, it can be dramatically enhanced in the vicinity of a biexciton ``Feshbach'' resonance \cite{Woutersresonant2007,takemura2014polaritonic,Bleu2020}. The use of the biexciton resonance to achieve strongly correlated polaritons was suggested in Ref.~\cite{Carusotto_2010}, but, to our knowledge, the polarization resolved quantum correlations have not been considered in this regime.
In fact, polarization-resolved second-order correlations have only been investigated numerically in the regime where the cross-circularly polarized polariton-polariton interactions are negligible  
\cite{Zhang2009,Bamba2011}.

Here, in light of the recent 
measurements of quantum correlations in exciton-polariton systems~\cite{delteil2019,MunozMatutano2019}, and new theoretical results for the interaction strengths \cite{Bleu2020}, we revisit the problem of a polariton pillar cavity coherently driven by a low-intensity polarized laser as shown schematically in Fig.~\ref{fig1}. 
In particular, we focus on the role of the different nonlinearities for co- and cross-circularly
polarized polaritons, $U_1$ and $U_2$, and the presence
of a birefringence which gives rise to an energy splitting between linearly polarized modes.
Our main original results are summarised below.

\paragraph*{Unconventional blockade --} We show that a substantial birefringence splitting combined with a circularly polarized drive allows one to realize the unconventional blockade in a single cavity. We derive the optimal conditions which give rise to vanishing second-order correlations between co-circularly polarized polaritons in the presence of cross-Kerr nonlinearity, i.e., when $U_2\neq0$.

\paragraph*{Conventional blockade --} 
The interplay between birefringence and nonlinearities also governs the correlations when the drive is linearly polarized.
In particular, we demonstrate 
that when the birefringence splitting is large compared to the other energy scales, the second-order correlations tend to be identical to that of a single-mode Kerr resonator with nonlinearity $U=(U_1+U_2)/2$.
We also compare our results with the recent correlation measurements \cite{delteil2019,MunozMatutano2019}, and we argue that these are likely to 
have probed the behavior close to the polariton Feshbach resonance where $|U_2|\gg U_1$. In particular, we find that the maximal antibunching observed in Ref.~\cite{delteil2019} is consistent with the analytical expressions for the polariton interactions derived in Ref.~\cite{Bleu2020}.

\paragraph*{Feshbach blockade --} We investigate the regime of resonantly enhanced $U_2$ using a two-channel model. Here, we introduce 
a new expression for the effective polariton-biexciton coupling, $g_{BL}$, which scales with the biexciton binding energy as $\sqrt{\epsilon_B^{XX}}$. Thus, one would expect an enhanced effect in monolayer semiconductors which support tighter bound biexcitons than quantum wells.  At resonance, we find that the relevant parameter quantifying the antibunching is $g_{BL}^2/\gamma\gamma_B$ with $\gamma, \gamma_B$ encoding polariton and non-radiative biexciton decays respectively. 

The paper is organized as follows. We first introduce the polariton system Hamiltonian in Sec.~\ref{Sec:Undriven}. In Sec.~\ref{Sec:Drive_Dissip}, we present the formalism used to describe the driven-dissipative system.  Then, we apply it to different driving scenarios. In Sec.~\ref{Sec:Circ}, we consider a circularly polarized drive, while the linearly polarized drive configuration is investigated in Sec.~\ref{Sec:Lin}. In Sec.~\ref{Sec:Feshbach}, we focus on the regime in the vicinity of the biexciton resonance using a two-channel model, and in Sec.~\ref{Sec:Conc} we conclude.
Additional details 
are provided in the appendices.

\section{System Hamiltonian}\label{Sec:Undriven}
In this section we introduce the model used to describe 
interacting lower polaritons confined in a box.
The motivations for working in the lower polariton (LP) subspace are twofold: (1) the LP spectral line is the easiest to precisely access experimentally; (2) from  the theory side, the composite electron-hole-photon nature of a single polariton has been shown to be less relevant for the LP for a broader range of Rabi couplings 
than is the case for the upper polariton \cite{Levinsen2019}. These reasons are ultimately related to the fact that the upper-polariton energy branch is closer to the electron-hole continuum than the lower polariton branch.

We first present the model for a perfectly symmetric cavity with degenerate polarization modes. Here, the use of Stokes operators allows us to highlight the fact that linear polarization is not a conserved quantity in the presence of interactions. Then, we introduce the birefringence splitting which can exist in realistic systems, and we discuss polariton-polariton interaction strengths.

\subsection{Perfect cavity}

\begin{figure}[tbp] 
   \includegraphics[width=\columnwidth]{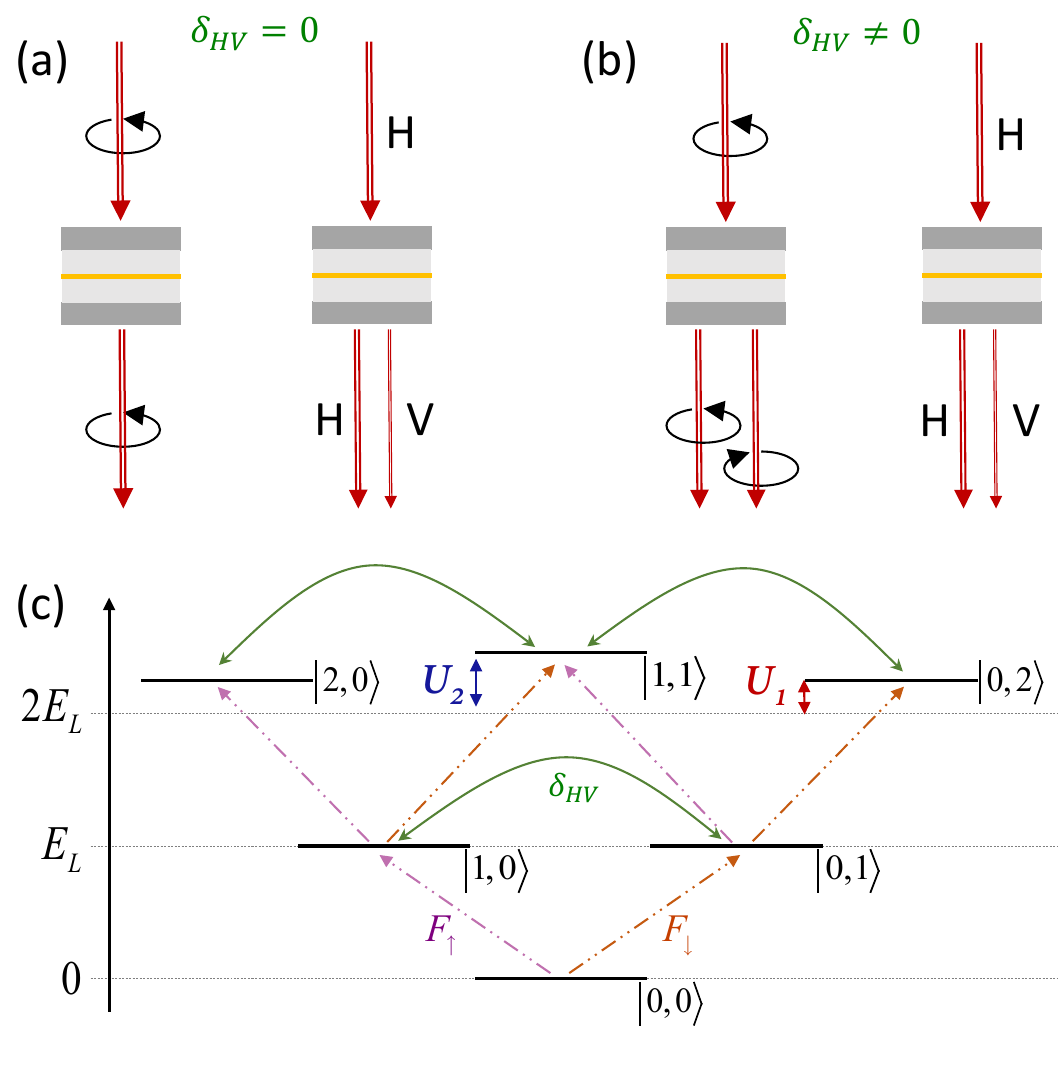}
\caption{Schematic representations of a perfectly symmetric cavity (a) and an anisotropic/birefringent cavity (b), 
which are resonantly driven with a circularly or linearly polarized laser. 
Polariton nonlinearities do not conserve linear (H or V) polarization, while birefringence does not conserve circular polarization.  (c) Sketch of the lowest energy levels for a perfectly symmetric cavity, where $\ket{n,m}$ corresponds to circularly polarized Fock states, with lower polariton energy $E_L$ and interaction energies $U_1$, $U_2$.  The presence of birefrigence leads to additional coupling ($\delta_{HV}\neq0$) between the levels, allowing the unconventional blockade to take place when the drive is circularly polarized.}
\label{fig1}
\end{figure}

To describe pairwise interacting lower polaritons in a pillar cavity, accounting for the polarization pseudospin, we use the following model Hamiltonian: 
\begin{eqnarray}\label{eq:Ham0}
   \hat{H}_0&=&\sum_{\sigma}E_L \hat{L}_{\sigma}^\dagger\hat{L}_{\sigma}+\sum_{\sigma,\sigma'}\frac{U_{\sigma\sigma'}}{2} \hat{L}_{\sigma}^\dagger\hat{L}_{\sigma'}^\dagger\hat{L}_{\sigma'}\hat{L}_{\sigma} .
\end{eqnarray}
 Here, $\sigma=\{\up,\dn\}$ encodes the \textit{circular} polarization degree of freedom, $\hat{L}_{\sigma}$ are the bosonic polariton annihilation operators, and $E_L$ is the lower polariton energy:
\begin{eqnarray}\label{eq:polEn}
E_L=E_X+ \frac{1}{2}\left(\delta-\sqrt{\delta^2+\Omega_R^2}\right),
\end{eqnarray}
where $E_X$ is the exciton energy, $\delta$ 
is the photon-exciton detuning and $\Omega_R$ is the Rabi splitting.
$U_{\sigma\sigma'}$ are the intra and inter-species interaction energies with $U_{\up\up}=U_{\dn\dn}\equiv U_1$ and
$U_{\up\dn}=U_{\dn\up}\equiv U_2$.
%
While we will focus on the polariton system here, we note that the Hamiltonian in Eq.~\eqref{eq:Ham0} is pretty generic. In the language of circuit QED, it represents two oscillators with degenerate mode frequencies, and both self- and cross-Kerr nonlinearities \cite{Kounalakis2018,Collodo2019}.

The eigenstates of the nonlinear Hamiltonian in Eq.~\eqref{eq:Ham0} correspond to Fock states with well-defined numbers of $\up$ and $\dn$ polaritons:
\begin{eqnarray}
  && \hat{H}_0\ket{n,m}= E_{nm}\ket{n,m} ,\\
   && E_{nm}=E_L\left(n+m\right)+\frac{U_1}{2}\left[n^2-n+m^2-m\right]+U_2 nm, \nonumber
\end{eqnarray}
where we have used 
$\hat{L}_{\up}\ket{n,m}=\sqrt{n}\ket{n-1,m}$ and $\hat{L}_{\dn}\ket{n,m}=\sqrt{m}\ket{n,m-1}$.
A sketch of the lowest energy levels is presented in Fig.~\ref{fig1}(c).
Importantly, the \textit{linearly} polarized Fock states are not eigenstates of the Hamiltonian \eqref{eq:Ham0}, except when $U_1=U_2$.
Hence, the Fock basis consisting of the circularly polarized eigenstates is the physically natural choice when accounting for polariton interactions. 

A useful way to visualize this important property is to introduce the Stokes operators for the polariton field in analogy with quantum optics \cite{Luis2002}:
%
%
\begin{equation}
\begin{aligned}
\label{eq:Stokes}
\hat{S}_0&=\hat{L}_{\up}^\dagger\hat{L}_{\up}+\hat{L}_{\dn}^\dagger\hat{L}_{\dn},\\
\hat{S}_1&=\hat{L}_{\up}^\dagger\hat{L}_{\dn}+\hat{L}_{\dn}^\dagger\hat{L}_{\up},\\
\hat{S}_2&=-i\left(\hat{L}_{\up}^\dagger\hat{L}_{\dn}-\hat{L}_{\dn}^\dagger\hat{L}_{\up}\right),\\
\hat{S}_3&=\hat{L}_{\up}^\dagger\hat{L}_{\up}-\hat{L}_{\dn}^\dagger\hat{L}_{\dn}. 
\end{aligned}
\end{equation}
Their expectation values $\langle\hat{S}_i\rangle$ correspond to the classical Stokes parameters that characterize the polarization of the polariton field. $\hat{S}_0$ encodes the total intensity, while $\hat{S}_j$ ($j=1,2,3$) are related to the degree of horizontal-vertical, diagonal-antidiagonal, and right-left circular polarization, respectively.
Formally, the introduction of these operators is analogous to the mapping of two independent harmonic oscillators to an angular momentum as introduced by Schwinger \cite{SchwingerAngular}.
Thus, the Stokes operators satisfy the commutation rules: 

\begin{subequations}\label{eq:Stokes_commutation}
\begin{align}
\left[\hat{S}_1,\hat{S}_2\right]=2i\hat{S}_3, &\left[\hat{S}_2,\hat{S}_3\right]=2i\hat{S}_1, \left[\hat{S}_3,\hat{S}_1\right]=2i\hat{S}_2, 
\\
&\left[\hat{S}_j,\hat{S}_0\right]=0,
\end{align}
\end{subequations}
and moreover we have $\hat{S}_1^2+\hat{S}_2^2+\hat{S}_3^2=\hat{S}_0\left(\hat{S}_0+2\right)$.

Rewritten in terms of the operators in Eq.~\eqref{eq:Stokes}, the Hamiltonian \eqref{eq:Ham0} reads:
\begin{equation}\label{eq:HamStokes}
   \hat{H}_0=\left(E_L-U_1\right)\hat{S}_0+\left( U_1+U_2\right)\hat{S}_0^2+\left( U_1-U_2\right)\hat{S}_3^2.
\end{equation}
We see that while both $\hat{S}_0$ and $\hat{S}_3$ commute with $\hat{H}_0$ (the total number of polaritons and the circular polarization are conserved), $\hat{S}_1$ and $\hat{S}_2$ do not, and thus the corresponding observables are not conserved quantities, as illustrated in Fig.~\ref{fig1}(a).

 
 \subsection{Birefringence}
In experiment, there can be 
a fine energy splitting between \textit{linearly} polarized cavity modes, which we refer to as horizontal (H) and vertical (V). This can, for instance, arise due to a residual birefringence of the sample or an imperfect cylindrical symmetry of the pillar cavity. Such a splitting can also be deliberately introduced and controlled by tuning the cavity shape \cite{Gayral1998,Gerhardt2019}. 

The fine polarization splitting is typically small with respect to the polariton Rabi splitting, $|\delta_{HV}|\ll \Omega_R$. Thus, in the following, we neglect its effect on $U_1$, $U_2$ and $E_L$.
In the present circular polarization basis, such a birefringence 
can then be modelled by introducing an additional term in the polariton Hamiltonian of the form:
\begin{equation}
  \hat{H}_{HV}=  \delta_{HV}(\hat{L}_{\up}^\dagger\hat{L}_{\dn}+\hat{L}_{\dn}^\dagger\hat{L}_{\up}).
\end{equation}
Within this approximation, the horizontal and vertical polaritons have the corresponding energies $E_L^H=E_L+\delta_{HV}$ and $E_L^V=E_L-\delta_{HV}$, respectively.

The total system Hamiltonian $\hat{H}_{syst}=\hat{H}_0+\hat{H}_{HV}$ is equivalent to two coupled resonators with Kerr and cross-Kerr nonlinearities~\cite{Kounalakis2018,Collodo2019}. In terms of Stokes operators, it reads
 \begin{eqnarray}\nonumber
   \hat{H}_{syst}&=&\left(E_L-U_1\right)\hat{S}_0+\delta_{HV}\hat{S}_1\\ 
   &&+\left( U_1+U_2\right)\hat{S}_0^2+\left( U_1-U_2\right)\hat{S}_3^2. \label{eq:HamStokes2}
\end{eqnarray}
We can easily see that $\delta_{HV}\neq0$ implies $[\hat{S}_3,\hat{H}_{syst}]\neq0$; hence, in the presence of birefringence, neither the linear nor the circular polarization degrees are conserved, as illustrated in Fig.~\ref{fig1}(b). 

We note though that the nonconservation of $\hat{S}_3$ and $\hat{S}_1,\hat{S}_2$ are not equivalent since the former is a one-body effect while the latter, being related to polariton-polariton interactions, is of two-body origin.

 \subsection{Polariton interactions} \label{subsec:Interactions}
 
For the exciton-polariton box we consider, the parameters $U_i$ are related to the polariton-polariton interaction strengths as $U_{i}=\alpha_i/\mathcal{A}$,
with $\mathcal{A}$ the polariton mode spatial area. 
The experimental estimation of the polariton interaction strengths was undertaken in several works (for example in refs. \cite{Vladimirova2010,Ferrier2011,Estrecho2019}) and has proven to be a subtle task with discrepancies among the reported values \cite{Estrecho2019}.
%
%

For cavities with embedded two-dimensional (2D) semiconductor layers, we recently introduced new analytical expressions to estimate the interaction strengths between lower polaritons~\cite{Bleu2020}: 
\begin{subequations}\label{eq:interactions}
\begin{eqnarray}\label{eq:alpha1}
  \alpha_1=\frac{4\pi \hbar^2 X^4}{\mathcal{N} m_X \ln\left(\frac{\epsilon_{B}^X}{2 |E_L-E_X|}\right)},
  \\ \label{eq:alpha2}
  \alpha_2=\frac{4\pi \hbar^2 X^4}{\mathcal{N} m_X \ln\left(\frac{\epsilon_{B}^{XX}}{2 |E_L-E_X|}\right)} .
\end{eqnarray}
\end{subequations}
Here, $\mathcal{N}$ is the number of layers (which also appears in the Rabi splitting $\Omega_R=\sqrt{\mathcal{N}}g_R$, with $g_R$ being the Rabi coupling of a monolayer), $X=(1+\delta/\sqrt{\delta^2+\Omega_R^2})^{1/2}/\sqrt{2}$ is the exciton Hopfield coefficient, $m_X$ is the exciton mass, and $\epsilon_{B}^X$ and $ \epsilon_{B}^{XX}$ are 
the exciton and biexciton binding energies, respectively. 

Equations \eqref{eq:interactions} have been derived within a scattering $T$ matrix approach~\cite{levinsen2015strongly} assuming tightly-bound structureless excitons, 
and are expected to be accurate when $2|E_L-E_X|\ll \epsilon_{B}^X$. The logarithm in the denominator is a remnant of the 2D nature of the polariton-polariton scattering~\footnote{Note that the polariton-polariton scattering can be treated as 2D, even in a box cavity, provided the area $\mathcal{A}$ is large compared to the exciton size.}. Crucially, this logarithm is finite, in contrast to the standard low-energy limit of 2D scattering of excitons. This is due to the Rabi splitting of the polariton modes and the extremely small photon mass; and therefore, can be seen as a remarkable consequence of the strong light-matter coupling~\cite{Bleu2020}. We note that a similar enhancement of polariton-electron interaction has been found recently within a microscopic theory involving electrons, holes, and photons~\cite{li2020enhanced,li2020theory}. We also stress that such an enhancement is absent in approaches based on the Born approximation \cite{Ciuti1998,tassone1999exciton}.

The expressions in Eq.~\eqref{eq:interactions} exhibit resonances when the argument of the logarithm tends to unity. Only the resonance for $\alpha_2$ is physical because of the well defined bound biexciton state. This resonance corresponds to the polariton Feshbach resonance proposed and experimentally probed in single quantum well planar cavities~\cite{Woutersresonant2007,takemura2014polaritonic,Deveaud2016}. An indirect signature of this resonance might also have been seen recently in a microcavity embedding a MoSe$_2$ monolayer \cite{Stepanov2021}.
The existence of this resonance implies that the magnitude and the sign of $\alpha_2/\alpha_1$ can be controlled experimentally in a given sample, via the photon-exciton detuning $\delta$. Using Eq.~\eqref{eq:polEn}, one can determine the critical detuning at which the resonance occurs: $\delta_c\simeq(\Omega_R^2-{\epsilon_{B}^{XX}}^2)/2\epsilon_{B}^{XX}$.
Finally, it is worth noticing that Eq.~\eqref{eq:interactions} implies that the condition $\alpha_1=\alpha_2$ is unrealistic for polaritons because it would require $\epsilon_{B}^X=\epsilon_{B}^{XX}$.

In the following, we keep $U_1$, $U_2$ as arbitrary model parameters since our results are applicable to any system effectively described by the Hamiltonian \eqref{eq:HamStokes2}. However, we use Eq.~\eqref{eq:interactions} for a quantitative comparison with the experiments involving polaritons \cite{MunozMatutano2019,delteil2019} in Sec.~\ref{subsec:Exp}.

\section{Driven system}\label{Sec:Drive_Dissip}
Having discussed the system Hamiltonian, we now include the resonant drive and dissipation. Within the rotating frame picture, we introduce the Lindblad equation used in the numerical calculations, and the effective wave function approach that we use for the analytical calculations. 

\subsection{Drive and dissipation}

A coherent drive can be introduced with the term 
\begin{eqnarray}
   \hat{H}_{drive}&=&\sum_{\sigma}\left(F_{\sigma}\hat{L}_{\sigma}e^{i \omega_p t}+F_{\sigma}^{*}\hat{L}_{\sigma}^{\dagger}e^{-i \omega_p t}\right),
\end{eqnarray}
with $\omega_p$ the frequency of the pump and $F_\sigma$ the amplitude of the drive in the $\sigma$ polarization.
We are interested in the case where the drive frequency is close to the polariton mode resonance. Thus, in the following we work in the rotating frame, where
the total Hamiltonian 
becomes  
\begin{eqnarray}\label{eq:Htot}
   \hat{H}&=&\hat{R} \left(\hat{H}_{syst}+\hat{H}_{drive}\right) \hat{R}^{\dagger}+i\hbar \dot{\hat{R}}\hat{R}^{\dagger},
\end{eqnarray}
with $\hat{R}=e^{i\omega_p t \left( \hat{L}_{\up}^\dagger\hat{L}_{\up}+\hat{L}_{\dn}^\dagger\hat{L}_{\dn}\right)}$. This yields:
\begin{align}\nonumber
   \hat{H}&=\sum_{\sigma}\left(\Delta \hat{L}_{\sigma}^\dagger\hat{L}_{\sigma}+F_{\sigma}\hat{L}_{\sigma}+F_{\sigma}^{*}\hat{L}_{\sigma}^{\dagger}\right) 
   \\
   &+\sum_{\sigma,\sigma'}\frac{U_{\sigma\sigma'}}{2} \hat{L}_{\sigma}^\dagger\hat{L}_{\sigma'}^\dagger\hat{L}_{\sigma'}\hat{L}_{\sigma} + \delta_{HV}(\hat{L}_{\up}^\dagger\hat{L}_{\dn}+\hat{L}_{\dn}^\dagger\hat{L}_{\up}), \label{eq:HtotKerr}
\end{align}
with the detuning $\Delta=E_L-\hbar\omega_p$. We consider the case $|\Delta|\ll \Omega_{R}$ which is a necessary condition to neglect the upper-polariton modes. 

We assume that the open-dissipative system density matrix obeys the following Lindblad (Markovian) equation:
\begin{align}\label{eq:Lindblad}
  \hbar \frac{\partial\hat{\rho}}{\partial t}  &=-i \left[ \hat{H},\hat{\rho}\right]+\gamma \sum_{\sigma} \left(   \hat{L}_{\sigma}\hat{\rho}\hat{L}_{\sigma}^{\dagger} - \frac{1}{2}\{\hat{L}_{\sigma}^{\dagger}\hat{L}_{\sigma},\hat{\rho}\} \right) ,
\end{align}
where $\gamma$ encodes the decay rate of polaritons from the cavity. It is related to the bare cavity photon decay $\gamma_c$ as $\gamma\simeq(1-X^2)\gamma_c$. Since we are mainly interested in phenomena related to the system itself, as described by the Hamiltonian $\hat{H}$, we assume this decay to be independent of the polarization. 
We note that accounting for different decay rates for H and V polarized modes would give rise to cross decoherence terms of the form $\hat{L}_{\sigma}\hat{\rho}\hat{L}_{\sigma'}^{\dagger}$ or $\hat{\rho}\hat{L}_{\sigma}^{\dagger}\hat{L}_{\sigma'} $ in Eq.~\eqref{eq:Lindblad}.

In the following we study the steady-state scenario  $\partial\hat{\rho}_{ss}/\partial t=0$ in the low-drive limit. The corresponding density matrix can be calculated by expanding it in terms of circularly polarized Fock states:
\begin{eqnarray}
\hat{\rho}_{ss} &=& \sum_{n,m,n',m'} \rho_{n,m,n',m'}\ket{n,m}\bra{n',m'} .
\end{eqnarray}
In the simulations, we truncate the Fock space such that $n+m\leq 6$, which is sufficient for the convergence of the results presented.
The expectation value of an operator $\hat{O}$ in the steady state is then defined as $\langle\hat{O}\rangle=\text{Tr}[\hat{\rho}_{ss}\hat{O}]$, with $\text{Tr}[\hat{\rho}_{ss}]=1$.

\subsection{Effective wave function approach}\label{subsec:WF}

Here, we present the wave function approach which we use to derive analytical results. Such an approach is typically well suited to investigate weakly driven systems \cite{Carmichael1991,Eleuch2008,bamba2011origin,Flayac2017,Huang2018}, and it can  provide an easier physical interpretation.
In this approach, the polarization-independent decay is introduced via an effective non-Hermitian Hamiltonian:
\begin{eqnarray} \label{eq:effHam}
\hat{H}_{\rm eff}=\hat{H}-i\frac{\gamma}{2} \sum_{\sigma} \hat{L}_{\sigma}^{\dagger}\hat{L}_{\sigma}.
\end{eqnarray}
The wave function can be expanded in the Fock basis as 
\begin{equation}\label{eq:WF}
    \ket{\psi}=\sum_{n,m}C_{nm}\ket{n,m},
\end{equation}
where $C_{nm}$ are complex time-dependent coefficients,
and the operator expectation values are $\langle\hat{O}\rangle=\bra{\psi}\hat{O}\ket{\psi}$.
We note that this wave function approach does not account for the terms $\hat{L}_{\sigma}\hat{\rho}\hat{L}_{\sigma}^{\dagger}$ present in the Lindblad equation. In principle, the results of the Lindblad equation can be reproduced within a Monte-Carlo wave function approach including quantum jumps \cite{Dalibard1992,Molmer:93}.
These are neglected here and instead the validity of the analytical results can be checked by comparing with the numerical results obtained from the Lindblad equation.

Since we are primarily interested in the first and second-order correlation functions in the low-drive limit, it is instructive to introduce a truncated ansatz where the Fock states in Eq.~\eqref{eq:WF} are limited to those basis states satisfying $n+m\leq2$.
By projecting the Schrödinger equation onto this subspace, we obtain the evolution equations for the $C_{nm}$ coefficients:
\begin{widetext}
\begin{subequations}\label{eq:evol}
\begin{eqnarray}
i  \hbar\dot{C}_{00}&=&C_{10} F_\up +C_{01} F_\dn, \\
i  \hbar\dot{C}_{10}&=&C_{00} F_\up^*+ \tilde{\Delta} C_{10}+F_\dn C_{11}+\sqrt{2}F_\up C_{20}+\delta_{HV}C_{01}, \label{eq:18b}\\
i  \hbar\dot{C}_{01}&=&C_{00} F_\dn^*+ \tilde{\Delta} C_{01}+F_\up C_{11}+\sqrt{2}F_\dn C_{02}+\delta_{HV}C_{10},\\
i  \hbar\dot{C}_{11}&=&C_{10} F_\dn^*+C_{01} F_\up^*+\left(2\tilde{\Delta}+U_2\right)C_{11}+\sqrt{2}\delta_{HV}\left(C_{20}+C_{02}\right),\\
i  \hbar\dot{C}_{20}&=&\sqrt{2}C_{10} F_\up^*+\left(2\tilde{\Delta}+U_1\right)C_{20}+\sqrt{2}\delta_{HV}C_{11},\\
i  \hbar\dot{C}_{02}&=&\sqrt{2}C_{01} F_\dn^*+\left(2\tilde{\Delta}+U_1\right)C_{02}+\sqrt{2}\delta_{HV}C_{11}.\label{eq:18f}
\end{eqnarray}
\end{subequations}
\end{widetext}
where $\tilde{\Delta}\equiv\Delta-i\gamma/2$.
In the following, we focus on the steady state solutions of the above system, i.e., we take $\dot{C}_{nm}=0$. Since we consider the low drive limit where $|F_{\sigma}|\ll \gamma$, in general, the coefficients obey the hierarchy $|C_{00}|\gg |C_{01}|,|C_{10}|\gg |C_{11}|,|C_{20}|,|C_{02}|$ and the normalization condition can be approximated as $1=\sum_{nm}|C_{nm}|^2\simeq |C_{00}|^2$. We therefore impose $C_{00}=1$ and solve Eqs.~\eqref{eq:18b}-\eqref{eq:18f} with the left hand side set to zero.

In the truncated basis, the average occupations $\bar{n}_\sigma= \langle \hat{L}_{\sigma}^\dagger\hat{L}_{\sigma}\rangle$ and the first-order coherences $\bar{n}_{\sigma,-\sigma}= \langle\hat{L}_{\sigma}^\dagger\hat{L}_{-\sigma}\rangle$ reduce to:
\begin{subequations}\label{eq:pop}
\begin{eqnarray}
\bar{n}_\up&=&|C_{10}|^2+|C_{11}|^2+2|C_{20}|^2,\\
\bar{n}_\dn&=&|C_{01}|^2+|C_{11}|^2+2|C_{02}|^2,\\
\bar{n}_{\up\dn}&=&C_{01}C_{10}^*+\sqrt{2}C_{02}C_{11}^*+\sqrt{2}C_{11}C_{20}^*,\\
\bar{n}_{\dn\up}&=&C_{10}C_{01}^*+\sqrt{2}C_{20}C_{11}^*+\sqrt{2}C_{11}C_{02}^*. 
\end{eqnarray}
\end{subequations}
To quantify the correlations between polaritons, we consider the normalized zero-time-delay second-order correlation functions:
$g_{\sigma\sigma'}^{(2)} =\langle\hat{L}_\sigma^\dagger\hat{L}_{\sigma'}^\dagger\hat{L}_{\sigma'}\hat{L}_\sigma\rangle / (\bar{n}_\sigma \bar{n}_{\sigma'})$ --- see Appendix \ref{Sec:correlations} for a general discussion of the correlation functions. These take the form:
\begin{subequations}\label{eq:g2}
\begin{eqnarray}\label{eq::WFg++}
g_{\up\up}^{(2)}&=&\frac{2|C_{20}|^2}{\bar{n}_\up^2}, \\
g_{\dn\dn}^{(2)}&=&\frac{2|C_{02}|^2}{\bar{n}_\dn^2} ,\\
g_{\up\dn}^{(2)}&=&g_{\dn\up}^{(2)}=\frac{|C_{11}|^2}{\bar{n}_\up\bar{n}_\dn}, \\
g_{tot}^{(2)}&=&\frac{2|C_{11}|^2+2|C_{20}|^2+2|C_{02}|^2}{(\bar{n}_\up+\bar{n}_\dn)^2}.
\end{eqnarray}
\end{subequations}
Here, $g_{tot}^{(2)}$ is related to the probability of finding two polaritons at the same time within the cavity, regardless of their polarization.
On the other hand, $g_{\sigma\sigma'}^{(2)}$ 
is related to the probability of finding two co- or cross-circularly polarized polaritons at the same time within the cavity, and the experimental measurement of this correlation function therefore requires polarization filtering at the output of the cavity. 
Equations~\eqref{eq:evol}-\eqref{eq:g2} form the basis for all analytic expressions in Secs.~\ref{Sec:Circ} and \ref{Sec:Lin}.

\section{Circularly polarized drive}\label{Sec:Circ}
Now that we have introduced our model and formalism, we proceed to present our results.
In this section, we consider the configuration where the coherent drive is circularly polarized, i.e., we take $F_\up$ to be real and non-zero while $F_\dn=0$. We first consider a cavity without birefringence, which we show is equivalent to the case of a single Kerr resonator. Including birefringence, we then demonstrate that the unconventional polariton blockade can be achieved in a single pillar cavity due to interference between different pathways to two-polariton states.

\subsection{No birefringence: Single Kerr resonator}
First, we consider a cavity in the absence of birefringence, i.e.,  we take $\delta_{HV}=0$. In this case, the $\{\up,\dn\}$ circular polarization subspaces decouple and therefore the states $\ket{n,m}$ with $m\neq 0$ remain unoccupied since $F_\dn=0$. The problem thus reduces to that of a single-mode Kerr resonator (in the $\up$ polarization), investigated previously in Ref.~\cite{Ferretti2012}. Solving the stationary condition in Eq.~\eqref{eq:evol} in the limit of a low drive, we find the leading order expressions $C_{10}\simeq -F_\up/\tilde\Delta$ and $C_{20}\simeq F_\up^2/[\sqrt2\tilde\Delta(\tilde\Delta+U_1/2)]$, while $C_{01}=C_{02}=0$. In this scenario, the only well defined second order correlation function is $g_{\up\up}^{(2)}=g_{tot}^{(2)}$, and its analytical expression in the low drive limit is:
\begin{eqnarray}\label{eq:g++nodHV1}
g_{\up\up}^{(2)}&=&\frac{\Delta^2+\gamma^2/4}{\left(\Delta+U_1/2\right)^2+\gamma^2/4}.
\end{eqnarray}
This result precisely matches that obtained in Ref.~\cite{Ferretti2012}. 
Its minimal value with respect to $\Delta$ is:
\begin{subequations}
\begin{eqnarray}\label{eq:mingpp}
\min[g_{\up\up}^{(2)}]&=&1 -\frac{\sqrt{U_1^4+4U_1^2\gamma^2}-U_1^2}{2\gamma^2},  \\ \Delta_{min}&=&\frac{\sqrt{U_{1}^4+4U_{1}^2\gamma^2}-U_{1}^2}{4 U_{1}}.
\end{eqnarray}
\end{subequations}
Thus, in the absence of birefringence, a strong antibunching requires a strong Kerr nonlinearity $|U_{1}|\gg\gamma$, in which case $\min[g_{\up\up}^{(2)}]\simeq\gamma^2/U_{1}^2$.  
\new{The opposite regime $|U_{1}|\ll\gamma$ gives a small antibunching with $\min[g_{\up\up}^{(2)}]\simeq1-|U_1|/\gamma$.}

\subsection{Unconventional polariton blockade}

In the presence of birefringence, the steady state solution of Eq.~\eqref{eq:evol} results in the following leading order behavior of the $C_{nm}$ coefficients:
\small
\begin{subequations}
\begin{align}
&C_{10}\simeq \frac{-\tilde{\Delta}F_\up}{\tilde{\Delta}^2-\delta_{HV}^2}, \\
&C_{01}\simeq\frac{\delta_{HV}F_\up}{\tilde{\Delta}^2-\delta_{HV}^2}, \\
%
&C_{11}\simeq \frac{-\delta_{HV}F_\up^2}{\tilde{\Delta}^2-\delta_{HV}^2} \frac{\tilde{\Delta}+U_1/4}{ (\tilde{\Delta}+U_1/2)(\tilde{\Delta}+U_2/2)-\delta_{HV}^2 }, 
\\
\label{eq:C20}
&C_{20}\simeq\frac{F_\up^2/\sqrt{2}}{\tilde{\Delta}^2\!-\!\delta_{HV}^2}\frac{\tilde\Delta}{\tilde\Delta\!+\!U_1/2}\!\frac{(\tilde\Delta\!+\!U_1/2)(\tilde\Delta\!+\!U_2/2)\!+\!\delta_{HV}^2U_1/4\tilde\Delta}{(\tilde\Delta+U_1/2)(\tilde\Delta+U_2/2)-\delta_{HV}^2},
\\
%
&C_{02}\simeq\frac{\delta_{HV}^2F_\up^2/\sqrt{2}}{\tilde{\Delta}^2-\delta_{HV}^2}\frac{\tilde\Delta+U_1/4}{\tilde\Delta+U_1/2}\frac1{(\tilde\Delta+U_1/2)(\tilde\Delta+U_2/2)-\delta_{HV}^2}.
%
\end{align}
\end{subequations}
\normalsize
%
As expected, the two circular polarization subspaces are now coupled and one gets $C_{nm}\neq0$ for $m\neq0$.
The populations and first order coherences to leading order in $F_\up$ then read:
\begin{subequations}
\begin{eqnarray}
&&\bar{n}_\up\simeq F_\up^2\abs{\frac{\tilde \Delta}{\tilde \Delta^2-\delta_{HV}^2}}^2,
\\
&&\bar{n}_\dn\simeq F_\up^2\frac{\delta_{HV}^2}{\abs*{\tilde \Delta^2-\delta_{HV}^2}^2},
\\
&&\bar{n}_{\up\dn}\simeq -F_\up^2\frac{\delta_{HV}\tilde\Delta^*}{\abs*{\tilde \Delta^2-\delta_{HV}^2}^2},
\\
&&\bar{n}_{\dn\up}\simeq\bar{n}_{\up\dn}^*.
\end{eqnarray}
\end{subequations}
Thus, we find that the birefringence gives rise to a cross circularly polarized polariton population $\bar{n}_\dn$ and to a complex coherence $\bar{n}_{\up\dn}$. This effect is purely linear and exists also in the absence of interactions. 

\begin{figure}[tbp] 
    \includegraphics[width=\linewidth]{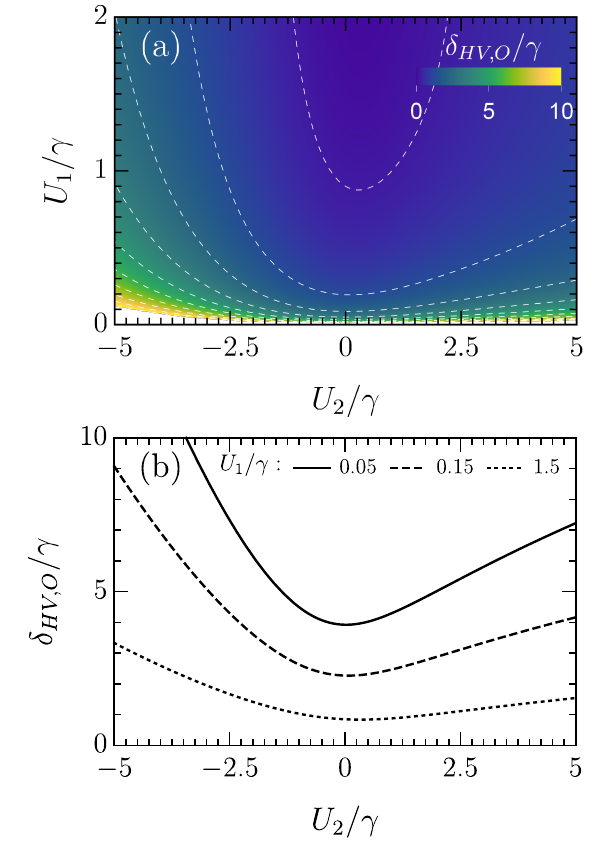}
\caption{(a) 2D colormap of the optimal birefringence coupling $\delta_{HV,O}$ versus the self and cross-Kerr nonlinearity strengths. The dashed white lines highlight contours where $\delta_{HV,O}/\gamma$ takes integer values. (b) Cross-sections of (a) for different values of $U_1/\gamma$.}
\label{fig:Opt2}
\end{figure}

\subsubsection{Optimal conditions}

The introduction of a birefringence coupling allows for different paths to populate the state $\ket{2,0}$, as illustrated in Fig.~\ref{fig1}(c). These can in turn interfere destructively and lead to a blockade of this state even when $U_1\ll\gamma$. 
This unconventional blockade mechanism was originally proposed for two coupled cavities in the absence of a cross-Kerr nonlinearity~\cite{Liew2010,bamba2011origin,Flayac2017}.  Here, we demonstrate that the effect survives in the presence of cross-Kerr nonlinearities. We derive the corresponding optimal conditions and the polarization resolved second-order correlations. 

From Eq.~\eqref{eq::WFg++}, we see that the antibunching should be maximal when $C_{20}\simeq0$. We find that optimal conditions satisfying $C_{20}=0$ --- which requires both the real and the imaginary part of the numerator in Eq.~\eqref{eq:C20} to vanish --- exist as long as $U_1\neq0$, and correspond to
\begin{subequations}\label{eq:Opt}
\begin{align}\label{eq:OptDet}
&\Delta_{O}=\frac{\pm\sqrt{3 \gamma ^2+U_1^2+U_2^2-U_1 U_2}-(U_1+U_2)}{6}, \\ \nonumber
&\delta_{HV,O}^2=\frac{\gamma ^2 (6 \Delta_O+U_1+U_2)}{2U_1}\\ 
& ~~~~~~~~~~~~ -\frac{2 \Delta_O (2 \Delta_O+U_1) (2 \Delta_O+U_2)}{2U_1}. \label{eq:Optdhv}
\end{align}
\end{subequations}
where the symbol $\pm$ in \eqref{eq:OptDet} corresponds to the sign of $U_1$. For polaritons, we have $U_1>0$ while $U_2$ can be either positive or negative. These expressions highlight the importance of an accurate estimation of the nonlinearities to predict the required optimal birefringence $\delta_{HV,O}$ and detuning $\Delta_{O}$. In the limit of small nonlinearities $\gamma\gg |U_i|$, these become independent of $U_2$ and reduce to
\begin{eqnarray}
\Delta_{O}\simeq \pm \frac{\gamma}{2\sqrt{3}}, ~~ \delta_{HV,O}\simeq \pm \left(\frac{4 \gamma^3}{3\sqrt{3}U_1}\right)^{1/2} .
\end{eqnarray}
Figure~\ref{fig:Opt2} illustrates the optimal birefringence splitting versus the interaction strengths.  We see that it increases when $U_1$ decreases and eventually diverges as $U_1\rightarrow 0$. One can also see an increase as a function of $|U_2|$ with a slight asymmetry between the $U_2>0$ and  $U_2<0$ sides, as shown in Fig.~\ref{fig:Opt2}(b). 


\subsubsection{Correlations}

Analytical expressions for the second-order coherences $g_{\sigma\sigma'}^{(2)}$ can be found in the limit $F_\up\rightarrow0$:
\begin{subequations}\label{eq:g2Circ} 
\begin{align}
\label{eq:g++Circ} 
   g^{(2)}_{\up\up} &\!=\!\left|\frac{\tilde\Delta^2-\delta_{HV}^2}{\tilde\Delta(\tilde\Delta\!+\!U_1/2)}\frac{(\tilde\Delta\!+\!U_1/2)(\tilde\Delta\!+\!U_2/2)\!+\!\delta_{HV}^2U_1/4\tilde\Delta}{(\tilde\Delta+U_1/2)(\tilde\Delta+U_2/2)-\delta_{HV}^2}
   \right|^2 \!\! ,\\ \label{eq:g--Circ}
   g^{(2)}_{\dn\dn} &\!=\!\left|\frac{\tilde\Delta+U_1/4}{\tilde\Delta+U_1/2}\frac{\tilde\Delta^2-\delta_{HV}^2}
   {(\tilde\Delta+U_1/2)(\tilde\Delta+U_2/2)-\delta_{HV}^2}\right|^2,\\
      \label{eq:g+-Circ}
   g^{(2)}_{\up\dn} &\!=\!\left|\frac{\tilde\Delta+U_1/4}{\tilde\Delta}\frac{\tilde\Delta^2-\delta_{HV}^2}{(\tilde\Delta+U_1/2)(\tilde\Delta+U_2/2)-\delta_{HV}^2}\right|^2.
\end{align}
\end{subequations}
We see that $g^{(2)}_{\up\up}$ vanishes for the optimal conditions given in Eqs.~\eqref{eq:OptDet} and \eqref{eq:Optdhv} and thus clearly exhibits unconventional antibunching. 
 In the special case $U_2=0$, Eq.~\eqref{eq:g++Circ} corresponds to the correlation function calculated numerically in Ref.~\cite{bamba2011origin}. In the limit of $\delta_{HV}\rightarrow 0$, Eq. $\eqref{eq:g++Circ}$ reduces to Eq.~\eqref{eq:g++nodHV1}, as it should.

From the above expressions, we can see that  $g^{(2)}_{\dn\dn}$ and  $ g^{(2)}_{\up\dn}$ become identical in the absence of self-Kerr nonlinearity ($U_1\rightarrow 0$). While optimal conditions do not exist for this latter case, non-trivial bunching or antibunching $g_{\sigma\sigma'}^{(2)}\neq1$ can still occur because of the cross-Kerr nonlinearity as soon as $\delta_{HV}\neq0$.
 
\begin{figure}[tbp] 
    \includegraphics[width=0.9\columnwidth]{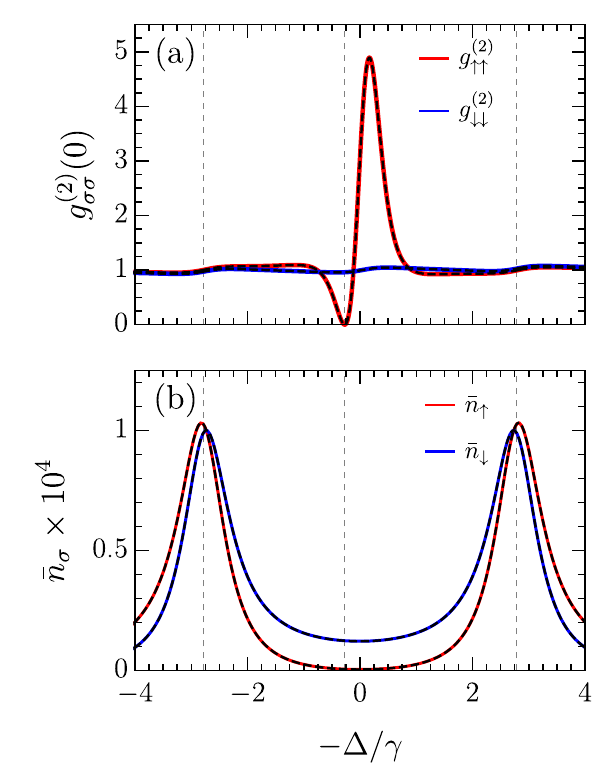}
\caption{Unconventional antibunching in the absence of cross-Kerr nonlinearity. (a) Second-order correlation functions for co-circular polaritons, and (b) average occupations $\bar{n}_\sigma$. 
The colored solid lines are calculated numerically from the steady state density matrix while the dashed-black lines represent the corresponding analytical results obtained within the wave function approach. The thin vertical dotted lines highlight $-\Delta_O/\gamma$ and $\pm\delta_{HV,O}/\gamma$ given by Eq.~\eqref{eq:Opt}. (Parameters: $ U_{1}=0.1 \gamma$, $U_{2}=0 \gamma$,  $\delta_{HV}=\delta_{HV,O}\simeq2.78\gamma$, $F_\up=0.01\gamma$)}
\label{fig:CircP1}
\end{figure}

\begin{figure}[tbp]
    \includegraphics[width=0.9\linewidth]{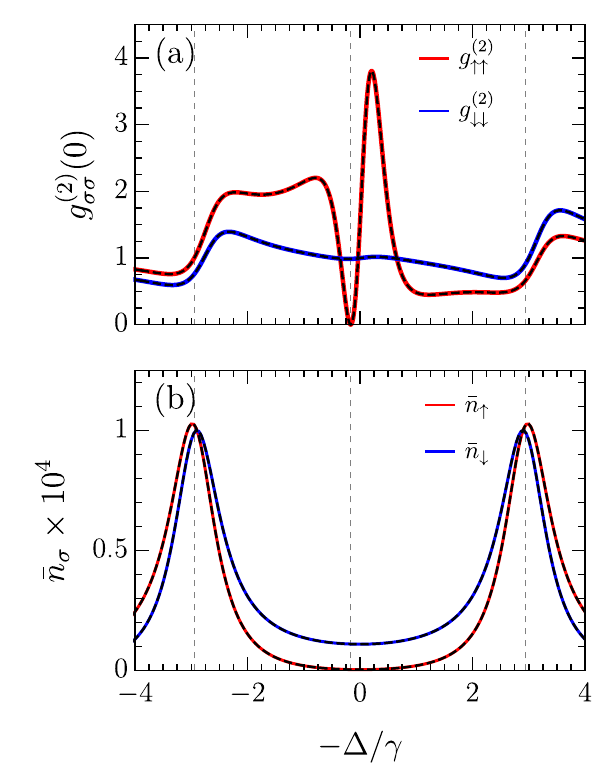}
\caption{Same as Fig.~\ref{fig:CircP1} in presence of cross-Kerr nonlinearity $U_{2}=0.8 \gamma$ which gives $\delta_{HV,O}\simeq2.94\gamma$.}
\label{fig:CircP2}
\end{figure}
 
In Figs.~\ref{fig:CircP1} and \ref{fig:CircP2}, we have plotted the second order correlations $g_{\sigma\sigma}^{(2)}$ and populations for the optimal $\delta_{HV,O}$ in the absence or presence of cross-Kerr nonlinearity. In both cases, $g_{\up\up}^{(2)}$ vanishes at the optimal detuning $\Delta_O$. It is worth noticing that $\Delta_O$ lies in between $\pm \delta_{HV,O}$, at a detuning where the populations shown in Figs.~\ref{fig:CircP1}(b) and \ref{fig:CircP2}(b) are dominated by $\bar{n}_{\dn}$. By comparing Figs.~\ref{fig:CircP1}(a) and \ref{fig:CircP2}(a), we can observe that cross-Kerr nonlinearities affect the form of the correlations as well as the exact values of $\Delta_O$ and $\delta_{HV,O}$. The fact that unconventional blockade survives in the regime where $|U_2|>U_1$ is particularly interesting for polariton systems since it corresponds to the interaction regime predicted in monolayer cavities~\cite{Bleu2020}.

\begin{figure}
    \includegraphics[width=0.9\linewidth]{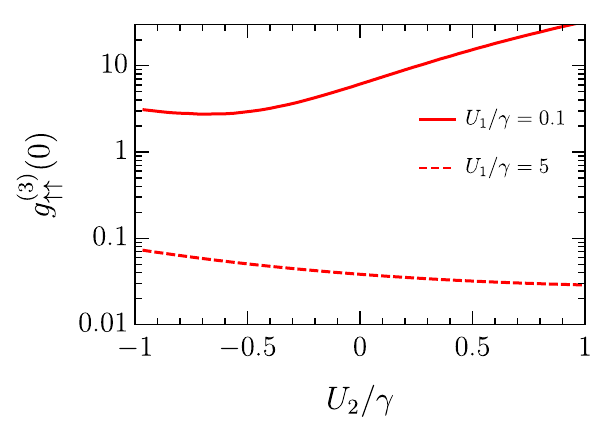}
\caption{Third-order correlation functions $g_{\up\up}^{(3)}$ at optimal conditions ($\Delta_O$, $\delta_{HV,O}$) for $ U_{1}=0.1 \gamma$ , and $ U_{1}=5 \gamma$ ($F_\up=0.01\gamma$).}
\label{fig:g3++}
\end{figure}

We conclude this section by emphasising that the above optimal conditions \eqref{eq:Opt} imply a blockade of the two-particle state $\ket{2,0}$ only. In contrast to the conventional Kerr blockade where the strong anharmonicity prohibits the occupation of any $n$-photon states with $n>1$, here, there is nothing that 
forbids the occupation of the states $\ket{n>2,0}$ when $U_1\ll\gamma$. To illustrate this point, we have plotted the third order correlations $g_{\up\up}^{(3)}$ calculated from the Lindblad equation for the optimal parameters (i.e., when $g_{\up\up}^{(2)}\simeq0$) for two values of $U_1/\gamma$ in Fig.~\ref{fig:g3++}. While its exact value depends on $U_2/\gamma$, we can see that $g_{\up\up}^{(3)}$ is always bunched when $U_1/\gamma=0.1$. Thus, in this configuration, the probability to occupy the state $\ket{3,0}$ is enhanced with respect to the case of a single driven resonator in the absence of nonlinearity. 
We note that such a bunching in higher-order coherences is a limitation of the unconventional mechanism for the generation of single photons, as noticed in Ref.~\cite{carreo2016criterion}.

\section{Linearly polarized drive}\label{Sec:Lin}

We now consider a horizontally (H) polarized drive with $F_\up=F_\dn\equiv F$ (i.e., symmetric driving), which corresponds to the experimental configuration of Refs.~\cite{delteil2019,MunozMatutano2019} in which quantum correlations of interacting polaritons were reported. 
Note that we can take $F$ to be real without loss of generality. 
While no unconventional blockade is expected here, the interplay between birefringence and nonlinearities still determines the correlations.
We first emphasize that $U_1\neq U_2$ necessarily leads to a small but non-zero cross-linear polarization population, as depicted in Fig.~\ref{fig1}. We then present analytical expressions for the second-order correlation functions and discuss three limiting cases: no birefringence, large birefringence, and weak nonlinearities. 
We conclude with a brief discussion of the correlations observed in the recent experiments~\cite{delteil2019,MunozMatutano2019}.

\subsection{Interaction-induced cross-polarization}

The averaged populations and first-order coherences up to 4th order in the symmetric drive $F$ read:
%
%
\begin{subequations} \label{eq:pop4th}
\begin{align}
\bar{n}_\up=&\bar{n}_\dn\simeq\frac{ F^2}{(\Delta+\delta_{HV})^2+\gamma^2/4}+F^4 f_1,\\
\bar{n}_{\up\dn}=&\bar{n}_{\dn\up}\simeq\frac{ F^2}{(\Delta+\delta_{HV})^2+\gamma^2/4}+F^4 f_2,
\end{align}
\end{subequations}
where $f_1$ and $f_2$ 
are real functions of the parameters $\{\gamma,\Delta,\delta_{HV},U_1,U_2\}$. 
In addition, the populations in the horizontally and vertically polarized modes are, respectively, $\bar{n}_H = \bar{n}_\up+\bar{n}_{\up\dn}$ and $\bar{n}_V = \bar{n}_\up-\bar{n}_{\up\dn}$ (see Appendix~\ref{Sec:correlations}).
Thus, at second order in $F$, we obtain $\bar{n}_V \simeq 0$ 
such that there is only a population in
the H-polarized driven mode. 
%
However, at higher orders this is no longer true, and Eq.~\eqref{eq:pop4th} highlights how the leading order corrections appear as $F^4$. 
Within the wave-function approach, the cross-polarization (V) population in the limit of weak driving is fully determined by the occupation of the 2-V polariton state, since it is generated by a two-polariton interaction term in the Hamiltonian (see Appendix~\ref{Appendix:Lin}).
In this case, the difference $f_1 -f_2$ yields the population: 
%
\begin{eqnarray}\label{eq:ny}
\bar{n}_{V,wf} &\simeq& \frac{16 F^4}{ 4(\Delta+\delta_{HV})^2+\gamma^2}\\ \nonumber
&&\times \frac{ \left(U_1-U_2\right)^2}{ \left(D_1 D_2-4\delta_{HV}^2\right)^2+\gamma^2\left( D_1^2+D_2^2+8\delta_{HV}^2\right)+\gamma^4 },
\end{eqnarray}
where we have introduced $D_i=(2\Delta+U_i)$.
Hence, we see that the non-zero population in the cross-linear polarization with respect to the drive is related to the non-equal nonlinearities $U_1\neq U_2$, as anticipated. 
In terms of the classical Stokes parameters $S_i\equiv\langle\hat{S}_i\rangle$, the emergence of a nonzero cross-polarization population can be interpreted as a slight \textit{depolarization} of the polariton field with respect to the coherent laser drive.
%
%
Specifically, one has $S_1^2+S_2^2+S_3^2<S_0^2$ as soon as $U_1\neq U_2$~\footnote{Note that, classically, a perfect polarization requires $S_1^2+S_2^2+S_3^2=S_0^2$, but this notion is more subtle in quantum optics because of the commutation rules of the Stokes operators \eqref{eq:Stokes_commutation} \cite{Luis2002,Goldberg2017}.}.

%

\begin{figure}[tbp] 
\centering
   \includegraphics[width=0.7\linewidth]{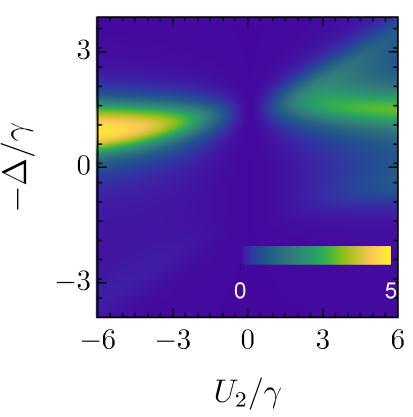}
\caption{
Interaction induced depolarization.  2D colormap of $\bar{n}_V\times10^8$  versus $U_2$ and detuning $-\Delta=\hbar\omega_p-E_L$. Parameters: $F=0.01\gamma$, $U_1=0.1\gamma$, $\delta_{HV}=1.5\gamma$.}
\label{fig4}
\end{figure}

Figure~\ref{fig4} displays the cross-polarization population $\bar{n}_V$ obtained numerically from the Lindblad equation for the weakly driven system, where we have used experimentally realistic parameters.    
We find that the analytic expression in  
Eq.~\eqref{eq:ny} accurately describes the results for the whole plotted parameter range provided we multiply $\bar{n}_{V,wf}$ by a factor of 2. Indeed, a perturbative expansion of the density matrix up to order $F^4$ yields $\bar{n}_{V}=2\bar{n}_{V,wf}$ (see Appendix~\ref{Appendix:Lin}).
This discrepancy between the density-matrix and the wave-function approaches is due to the fact that the 2-V polariton state can decay into the 1-V polariton state 
via the terms $\hat{L}_{\sigma}\hat{\rho}\hat{L}_{\sigma}^{\dagger}$ in the Lindblad equation, which are not present in the effective non-Hermitian Hamiltonian, Eq.~\eqref{eq:effHam}. 
In the case of a circularly polarized drive, such ``quantum-jump'' terms are only relevant at higher order in the driving strength, since the full set of number states can be populated by the Hamiltonian.

 
\subsection{Second-order correlations}

As we did in the previous section for the circularly polarized drive, we now calculate the second-order correlations at zero time delay. 
Note that there is no unconventional blockade in the case of a linearly polarized drive (in the sense that there are no optimal conditions for it),  
but the polariton-polariton interactions can still give rise to non-trivial correlations.

Since the H-polarized drive is symmetric with respect to the $\{\up,\dn\}$ subspaces, we have the relations
\begin{subequations}\label{eq:g2sym}
\begin{eqnarray}
g_{\up\up}^{(2)}&=&g_{\dn\dn}^{(2)}=g_{\sigma\sigma}^{(2)},\\
\label{eq:gtotLin}
g_{tot}^{(2)}&=&\frac{1}{2}\left(g_{\sigma\sigma}^{(2)}+g_{\up\dn}^{(2)}\right).
\end{eqnarray}
\end{subequations}
%
In the low-drive limit $F/\gamma \ll 1$, the polarization-resolved correlation functions have the following analytical expressions:
%

\begin{subequations}\label{eq:g+++-Lin}
\begin{align}
\label{eq:g++Lin}
g_{\sigma\sigma}^{(2)}&=\frac{\left(4(\Delta+\delta_{HV})^2+\gamma ^2\right) \left((D_2-2\delta_{HV})^2+\gamma ^2\right)}{ \left(D_1 D_2-4\delta_{HV}^2\right)^2+\gamma^2\left( D_1^2+D_2^2+8\delta_{HV}^2\right)+\gamma^4 },\\ \label{eq:g+-Lin}
g_{\up\dn}^{(2)}&=\frac{\left(4 (\Delta+\delta_{HV})^2+\gamma ^2\right) \left((D_1-2\delta_{HV})^2+\gamma ^2\right)}{ \left(D_1 D_2-4\delta_{HV}^2\right)^2+\gamma^2\left( D_1^2+D_2^2+8\delta_{HV}^2\right)+\gamma^4 }.
\end{align}
\end{subequations}
We note that Eq.~\eqref{eq:g++Lin} matches the result of Ref.~\cite{Ferreti2010} in the special case where $\Delta=U_2=0$.
While Eqs.~\eqref{eq:g++Lin} and \eqref{eq:g+-Lin} are quite cumbersome, 
their ratio takes the simple form:
\begin{eqnarray}\label{eq:ratio}
\frac{g_{\up\dn}^{(2)}}{g_{\up\up}^{(2)}}=\frac{(2 \Delta-2\delta_{HV}+U_1)^2+\gamma ^2}{(2 \Delta -2\delta_{HV} +U_2)^2+\gamma ^2},
\end{eqnarray}
 which we see approaches unity when $U_2\simeq U_1$, or when $|\Delta - \delta_{HV}|\gg |U_i|$.
In addition, the ratio \eqref{eq:ratio} and the form of Eq.~\eqref{eq:gtotLin} imply that one cannot determine which of the nonlinearities ($U_1$ or $U_2$) is dominant from a measurement 
of the total intensity correlation alone. However, it is possible to extract the relative strength of the nonlinearities from polarization-resolved measurements, as we discuss further below.

One can similarly obtain the correlation functions in the linearly rather than the circularly polarized basis, but the derivation is more involved since we must account for quantum-jump processes when calculating the V-polariton states. We therefore relegate these to Appendix~\ref{Appendix:Lin}. 

In the following, we discuss several limiting regimes.

\subsubsection{No birefringence}

In the absence of birefringence ($\delta_{HV}=0$), the problem is equivalent to two equally driven and uncoupled oscillators with self- and cross-Kerr nonlinearities. Equation \eqref{eq:g+++-Lin} 
then reduces to:
\begin{subequations}\label{eq:glinnodHV}
\begin{eqnarray}\label{eq:g++nodHV}
g_{\sigma\sigma}^{(2)}&=&\frac{\Delta^2+\gamma^2/4}{\left(\Delta+U_1/2\right)^2+\gamma^2/4},\\\label{eq:g+-nodHV}
g_{\up\dn}^{(2)}&=&\frac{\Delta^2+\gamma^2/4}{\left(\Delta+U_2/2\right)^2+\gamma^2/4}.
\end{eqnarray}
\end{subequations}
We can see that $g_{\sigma\sigma'}^{(2)}$ has the same form as that obtained for a single-mode Kerr resonator in Eq.~\eqref{eq:g++nodHV1}.
Hence, a strong antibunching for co- or cross-circularly polarized polaritons requires strong nonlinearities $|U_{\sigma\sigma'}|\gg\gamma$.

The physical meaning of these expressions is particularly transparent for $\Delta=0$, in which case the relative strength of the nonlinearities completely determines whether it is more favorable to find two co-polarized 
$(g_{\sigma\sigma}^{(2)}>g_{\up\dn}^{(2)})$ or two cross-polarized 
$(g_{\sigma\sigma}^{(2)}<g_{\up\dn}^{(2)})$ polaritons.

\subsubsection{Large birefringence splitting}\label{subsec:Largedhv}
 
In the opposite limit, where the birefringence splitting is much larger than the other energy scales, it is insightful to express the correlations in terms of the laser detuning from the driven H-polarized mode: $\Delta_H=\Delta+\delta_{HV}$. 
 In this case, the second-order correlations are independent of $\delta_{HV}$ if $|\delta_{HV}|\gg |\Delta_H|$, and Eq.~\eqref{eq:g+++-Lin} simplifies to:
 \begin{eqnarray}\label{eq:g++largedHV}
g_{\sigma\sigma}^{(2)}\simeq g_{\up\dn}^{(2)}\simeq\frac{\Delta_H^2+\gamma^2/4}{\left(\Delta_H+U/2\right)^2+\gamma^2/4},
\end{eqnarray}
with $U=(U_1+U_2)/2$.
This expression is identical to Eq.~\eqref{eq:g++nodHV1}, with the average non-linearity $U$ replacing $U_1$ and $\Delta_H$ replacing $\Delta$. 
%
Moreover, the total correlation function is also approximately 
$g_{tot}^{(2)} \simeq g_{\sigma\sigma}^{(2)}\simeq g_{\up\dn}^{(2)}$. 
Thus, the case of large birefringence resembles a single (H-polarized) Kerr resonator with $U$ nonlinearity, where we can neglect the cross-polarization (V) population. We note that the regime $|\delta_{HV}|\gg\gamma$ seems achievable in high finesse fiber cavities \cite{Besga2015} and can be controlled artificially by tuning the cavity shape \cite{Gayral1998,Gerhardt2019}. There is also the prospect of enhancing the nonlinearity $U$ via the use of a Feshbach resonance in the interspecies interaction $U_2$, as we discuss in Sec.~\ref{Sec:Feshbach}.


\subsubsection{Weak nonlinearities}

\begin{figure}[tbp]
   \includegraphics[width=0.9\linewidth]{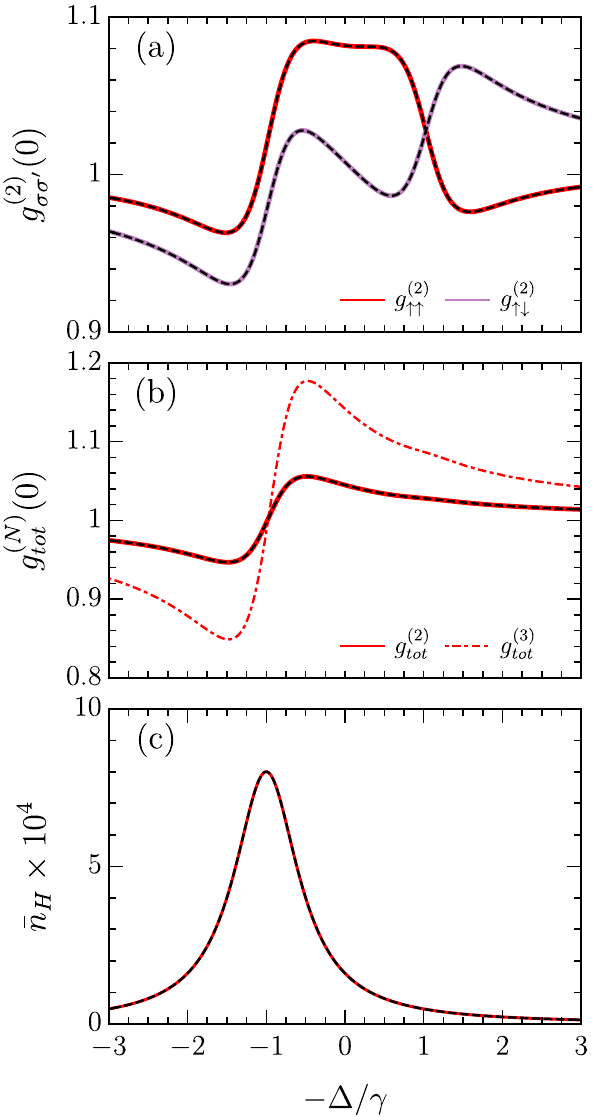}
\caption{Correlation functions versus detuning $\Delta$ under weak H drive for the parameters: $U_1=0.01\gamma$, $U_2=0.1\gamma$, $\delta_{HV}=-\gamma$, $F=0.01\gamma$.   (a) Second-order auto- and cross-correlation functions $g_{\sigma\sigma'}^{(2)}$. (b) Second- and third-order total intensity correlations. (c) Average population $\bar{n}_H$.  
The solid colored lines and dot-dashed red lines correspond to the numerical results from the Lindblad equation while the dashed-black lines correspond to the analytical results (Eqs.~\eqref{eq:g+++-Lin}, \eqref{eq:gtotLin}). }
\label{fig:g2lin} 
\end{figure}

The birefringence splitting can also be similar in size to the linewidth such that $|\delta_{HV}|\sim \gamma$, which is the case in the experiments of Refs.~\cite{delteil2019,MunozMatutano2019}. Therefore
 none of the above limiting cases is strictly applicable. Nevertheless, in this case
 it is instructive to consider the regime of small nonlinearities $U_1,|U_2|\ll\gamma,|\delta_{HV}|$, for which Eq.~\eqref{eq:gtotLin} can be Taylor expanded to give:
\begin{eqnarray}
g_{tot}^{(2)}&\simeq& 1-\frac{2 \Delta_H }{\gamma ^2+4 \Delta_H^2} (U_1+U_2). \label{eq:gtot_SmallNonLin}
\end{eqnarray}
Assuming constant $U_1$ and $U_2$, Eq.~\eqref{eq:gtot_SmallNonLin} is minimized for the conditions $\Delta_H=\pm\gamma/2$ (with $+$ if $(U_1+U_2)>0$, and $-$ if $(U_1+U_2)<0$). The corresponding minimal value then reads
\begin{eqnarray}\label{eq:gtotMin}
\min[g_{tot}^{(2)}]=1-\frac{|U_1+U_2|}{2\gamma}.
\end{eqnarray}
As an example, in Fig.~\ref{fig:g2lin}, we have plotted the correlations for the regime where $\gamma\sim|\delta_{HV}|\gg U_2\gg U_1$. Panel (a) illustrates how $g_{\up\up}^{(2)}$ and $g_{\up\dn}^{(2)}$ can significantly differ from each other when $U_1\neq U_2$. In addition, their shapes are  qualitatively different from $g_{tot}^{(2)}$ plotted in Fig.~\ref{fig:g2lin}(b). We also observe that the largest antibunching in $g_{tot}^{(2)}$ (corresponding to Eq.~\eqref{eq:gtotMin}) 
is located at a detuning where the population $\bar{n}_H$ shown in Fig.~\ref{fig:g2lin}(c) is substantial. This feature is similar to the conventional antibunching in a single Kerr resonator, and differs from the unconventional case displayed in Figs.~\ref{fig:CircP1} and \ref{fig:CircP2}.
Finally, we see that the third-order correlation function $g_{tot}^{(3)}$ in panel (b) 
also exhibits antibunching, which confirms that the total-intensity antibunching is related to the substantial strength of $U_1+U_2$ and is therefore of ``conventional'' origin.

\subsection{Comparison with experiment}\label{subsec:Exp}
In the recent experiments \cite{delteil2019,MunozMatutano2019}, the total intensity correlations have been measured in a fiber cavity with $|\delta_{HV}|\sim\gamma$, and the reported nonlinearity was small compared to the linewidth $U\ll\gamma$. Thus one can use the results presented in the above subsection to reinterpret the measurements.

For example, in Ref.~\cite{delteil2019}, the maximal antibunching reported is $\min[g_{exp}^{(2)}]\simeq 0.95$ for a linewidth $\gamma_{exp}=\hbar/\tau_p\simeq 60\mu$eV (i.e., a polariton lifetime $\tau_p\simeq11ps$).
Furthermore, in the experiment, the maximal antibunching was observed for $\Delta_H >0$, which according to Eq.~\eqref{eq:gtot_SmallNonLin} requires $U_1+U_2>0$. Comparing with Eq.~\eqref{eq:gtotMin}, one then obtains $(U_1+U_2)^{exp}\simeq 6 \mu \text{eV}$.
%
This value is too large to be explained by an estimation of polariton interactions within the Born approximation, where $U_1^{Born}\simeq 6\epsilon_B^Xa_B^2X^4/\mathcal{A}$ and  $U_2^{Born}\simeq 0$ \cite{tassone1999exciton}, with $a_B$ the Bohr radius. 
Indeed, the use of the experimental parameters $\Omega_R=3.5$meV, $\delta=0.72$meV ($X^2\simeq0.6$), $\mathcal{A}=\pi \mu$m$^2$ \cite{delteil2019}, and $a_{B,\text{GaAs}}\simeq10$nm, $\epsilon_{B,\text{GaAs}}^{X}\simeq10$meV gives $U_1^{Born}\simeq 0.69\mu$eV. 

Going beyond the Born approximation, we instead compare with the experimental results using the analytic polariton-polariton interactions in Eq.~\eqref{eq:interactions} which takes higher-order interaction processes into account~\cite{Bleu2020}.  Using $m_{X,\text{GaAs}}=0.63m_0$ \cite{NAKWASKI19951} (with $m_0$ the free electron mass), and the experimental parameters, we obtain $U_1^{exp}\simeq 0.14\mu$eV from Eq.~\eqref{eq:alpha1}. This is clearly much too small to explain the experimental result of $(U_1+U_2)^{exp}\simeq 6 \mu \text{eV}$. Instead, the experiment suggests that the singlet interaction is much larger than the triplet one, $U_2^{exp}\simeq42 \times U_1^{exp}$. While this is not predicted within the Born approximation~\cite{tassone1999exciton}, it is consistent with our Eq.~\eqref{eq:alpha2}. Indeed, within logarithmic accuracy, one can use the ratio $U_2/U_1$ to estimate the associated biexciton binding energy as $\epsilon_{B}^{XX}=2|E_L-E_X|\left(\frac{\epsilon_{B}^{X}}{2|E_L-E_X|}\right)^{1/42}\simeq 2.94$meV, which is a reasonable value for  biexcitons in InGaAs quantum wells \cite{takemura2014polaritonic,Borri1999}.

\new{
The potentially crucial role played by polarization-dependent interactions in 
the measured antibunching~\cite{delteil2019} calls for additional polarization resolved experiments. 
A natural extension of the recent measurements would be to add a circular polarizer between the cavity output and the Hanbury Brown-Twiss setup, thus allowing access to $g_{\sigma\sigma}^{(2)}$. When the drive is linearly polarized, the relations given in Eq.~\eqref{eq:g2sym} imply that such a measurement, combined with one of $g_{tot}^{(2)}$, is sufficient to deduce $g^{(2)}_{\up\dn}$. We also note that a measurement of the cross-linear polarization population [Eq.~\eqref{eq:ny}] can be used to unveil the difference 
between $U_1$ and $U_2$. Thus, a direct test of 
our predictions seems within reach.

Our findings} suggest that the two recent experiments have probed a regime close to the polariton Feshbach resonance where $U_2$ is expected to diverge.
In the vicinity of the resonance, the effective nonlinearity between cross-circularly polarized polaritons can vary extremely quickly as a function of the polariton energy~\cite{Bleu2020,Woutersresonant2007} and the assumption of a constant $U_2$ is not accurate. In the following section, we therefore focus on 
this interesting regime. 

\section{Feshbach regime}\label{Sec:Feshbach}

In the coupled Kerr-resonators model, the interaction parameters $U_1$ and $U_2$ are assumed to be constant. In the polariton system, this assumption is accurate as long as the driven polariton mode is far from the ``Feshbach'' resonance condition,
in which case the variation of the nonlinearities for the range of $\Delta$ probed is negligible.
However, in the vicinity of the resonance, $U_2$ varies rapidly 
and this assumption is no longer appropriate. Here, we introduce a two-channel model that explicitly includes the biexciton state, allowing us to investigate this regime more accurately. Focusing on the horizontally polarized drive case, we derive the correlation functions in the low driving limit.

\subsection{Model}

The Feshbach resonance occurs when the energy of two lower polaritons of opposite polarization is close to the energy $E_{XX}$ of a bound biexciton state~\cite{Woutersresonant2007}.
In the vicinity of the resonance, one can model the system with an effective two-channel Hamiltonian encoding a coupling with a biexciton state:
\begin{eqnarray}\label{eq:HamB}\nonumber
   \hat{H}_{0}&=&\sum_{\sigma}E_L \hat{L}_{\sigma}^\dagger\hat{L}_{\sigma}+\sum_{\sigma}\frac{U_{\sigma\sigma}}{2} \hat{L}_{\sigma}^\dagger\hat{L}_{\sigma}^\dagger\hat{L}_{\sigma}\hat{L}_{\sigma}\\
   && + E_{XX} \hat{B}^\dagger\hat{B} + g_{BL}\left(\hat{B}^\dagger\hat{L}_{\up}\hat{L}_{\dn}+\hat{L}_{\dn}^\dagger\hat{L}_{\up}^\dagger\hat{B}\right),
\end{eqnarray}
where $E_{XX}=2E_X-\epsilon_B^{XX}$ is the biexciton energy, $\hat B$ is a biexciton operator, and $g_{BL}$ is the effective biexciton-polariton coupling strength. We determine the strength of this coupling by requiring that it yields the correct energy-dependent interspin nonlinearity \eqref{eq:alpha2} close to the resonance. This gives
%
%
\begin{equation}\label{eq:coupling}
g_{BL}\simeq X^2\left(\frac{4\pi\hbar^2\epsilon_B^{XX}}{\mathcal{N}m_X\mathcal{A}}\right)^{\frac{1}{2}},
\end{equation}
which differs from that used in Ref.~\cite{Carusotto_2010} in several important ways.
In particular, we note that the coupling increases with the biexciton binding energy as $\sqrt{\epsilon_{B}^{XX}}$, and it only depends on the Rabi splitting $\Omega_R$ through the Hopfield coefficient.


It is worth noticing that the Hamiltonian \eqref{eq:HamB} does not fulfill the same conservation laws as the coupled Kerr oscillators model. In contrast to Eq.~\eqref{eq:HamStokes2}, Eq.~\eqref{eq:HamB} cannot be written solely in terms of the Stokes operators because of the terms $\hat{B}^\dagger\hat{L}_{\up}\hat{L}_{\dn}$. Also, the polariton number is not conserved here and the relevant conserved number satisfying  $[\hat{H}_{syst},\hat{N}]=0$ is $\hat{N}= \hat{L}_{\up}^\dagger\hat{L}_{\up}+\hat{L}_{\dn}^\dagger\hat{L}_{\dn}+2\hat{B}^\dagger\hat{B}$, where the factor of 2 originates from the fact that two polaritons are required to create one biexciton.

\begin{figure}[tbp]
   \includegraphics[width=\linewidth]{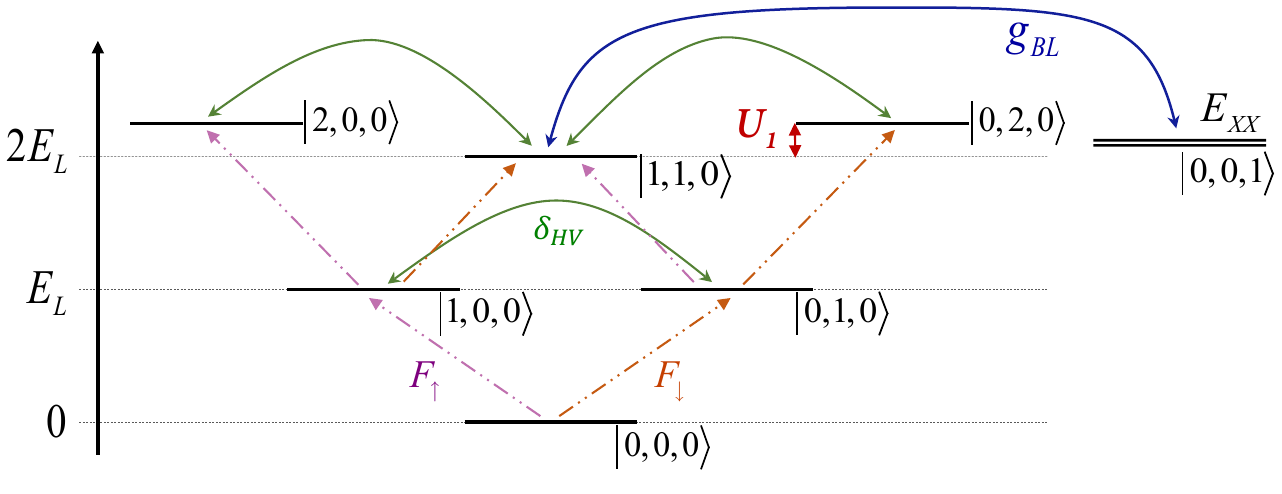}
\caption{Sketch of the first energy levels $\ket{n,m,o}$ in the two-channel model. The polariton-biexciton coupling $g_{BL}$ results in an effective nonlinearity between cross-circularly polarized polaritons. Note that the biexciton state is not directly excited by the laser drive.}
\label{fig:sketch2} 
\end{figure}

The birefringence and the external drive can be introduced in the same manner as in Sections \ref{Sec:Undriven} and \ref{Sec:Drive_Dissip}, where we note that the coherent drive $F_\sigma$ injects polaritons, not biexcitons, as illustrated in Fig. \ref{fig:sketch2}. Using the unitary transformation $\hat{R}=e^{i\omega_p t \hat{N}}$ in Eq.~\eqref{eq:Htot}, we then obtain the complete Hamiltonian in the rotating frame
\begin{eqnarray}\nonumber
   \hat{H}&=&\sum_{\sigma}\left(\Delta \hat{L}_{\sigma}^\dagger\hat{L}_{\sigma}+\frac{U_{\sigma\sigma}}{2} \hat{L}_{\sigma}^\dagger\hat{L}_{\sigma}^\dagger\hat{L}_{\sigma}\hat{L}_{\sigma}+F_{\sigma}\hat{L}_{\sigma}+F_{\sigma}^{*}\hat{L}_{\sigma}^{\dagger}\right)\\ 
   & &+ \Delta_B \hat{B}^\dagger\hat{B} + g_{BL}\left(\hat{B}^\dagger\hat{L}_{\up}\hat{L}_{\dn}+\hat{L}_{\up}^\dagger\hat{L}_{\dn}^\dagger\hat{B}\right)+ \hat{H}_{HV}. \label{eq:Ham_2chann}
\end{eqnarray}
Here, there are two relevant detunings: in addition to the detuning from the polariton mode, $\Delta=E_L-\hbar\omega_p$, we now also have the detuning from the Feshbach resonance, $\Delta_B=E_{XX}-2\hbar\omega_p$. 

As in the previous sections, we use a master equation to calculate the steady state density matrix numerically, and an effective wave function approach to obtain analytical results which are accurate in the low-drive limit. The \textit{non-radiative} decay of the biexciton excitations is introduced using a phenomenological parameter $\gamma_B$. Additional details are provided in Appendix~\ref{Sec:2chann}. 

In the following, we focus on the symmetric (H-polarized) drive with $F_\up=F_\dn=F$ for which the relations \eqref{eq:g2sym} are satisfied. The complete analytical expressions for the correlation functions
are provided in Appendix~\ref{Sec:g2_2chann}. Importantly, in the limit of vanishing biexciton decay $\gamma_B/\gamma\rightarrow 0$, one finds a one-to-one correspondence between the analytical expressions for the correlation functions obtained in Sections \ref{Sec:Circ} and \ref{Sec:Lin} 
and the present two-channel model upon the replacement $U_2\leftrightarrow -\frac{g_{BL}^2}{\Delta_B}$. 


\subsection{Behavior at resonance}

We first consider 
what happens when the driven polariton mode is tuned exactly on resonance: $2E_L^H
=E_{XX}$ (i.e., $\Delta_B=2\Delta_H$).
Then, taking the limit $\Delta_H\rightarrow0$ corresponding to a resonant pump, the correlations can be expressed as
%
%
\begin{subequations}\label{eq:g2Lin_res}
\begin{eqnarray}\label{eq:g++Lin_res}
g_{\sigma\sigma,res}^{(2)}&=& \frac{\gamma ^2 \left(16  \delta_{HV}^2+\gamma ^2 (1+2 x)^2\right)}{\gamma ^2 h+\left(2 \delta_{HV} U_1-\gamma ^2\right)^2},
\\ \label{eq:g+-Lin_res}
g_{\up\dn,res}^{(2)}&=&\frac{\gamma ^2 \left(\gamma ^2+(U_1-4\delta_{HV})^2\right)}{\gamma ^2 h+\left(2 \delta_{HV} U_1-\gamma ^2\right)^2},
\end{eqnarray}
\end{subequations}
where we have introduced $x=g_{BL}^2/\gamma\gamma_B$ and 
\begin{eqnarray}\nonumber
h&=&4 \left(1+x\right) \left(4 \delta_{HV}^2+x \left(\gamma ^2+(2
   \delta_{HV}-U_1)^2\right)\right)+U_1^2.
\end{eqnarray}
We see that when $\gamma_B\rightarrow 0$, $x \rightarrow \infty$ and thus $g_{\up\dn,res}^{(2)}\rightarrow 0$, as it should.

In the limit of vanishing birefringence $\delta_{HV}\rightarrow 0$, Eq.~\eqref{eq:g2Lin_res} simplifies to
\begin{subequations}\label{eq:gresonanceNodhV}
\begin{eqnarray}
g_{\sigma\sigma,res}^{(2)}&\simeq&\frac{1}{1+U_1^2/\gamma^2}, \\
g_{\up\dn,res}^{(2)}&\simeq &\frac{1}{(2g_{BL}^2/\gamma\gamma_B+1)^2}.
\end{eqnarray}
\end{subequations}
Hence, in this situation $g_{\sigma\sigma}^{(2)}$ is unaffected by the biexciton resonance in contrast to $g_{\up\dn}^{(2)}$ which is completely determined by it. This mirrors the corresponding result in the absence of a resonance, Eq.~\eqref{eq:glinnodHV}.

\begin{figure}[tbp]
   \includegraphics[width=0.9\linewidth]{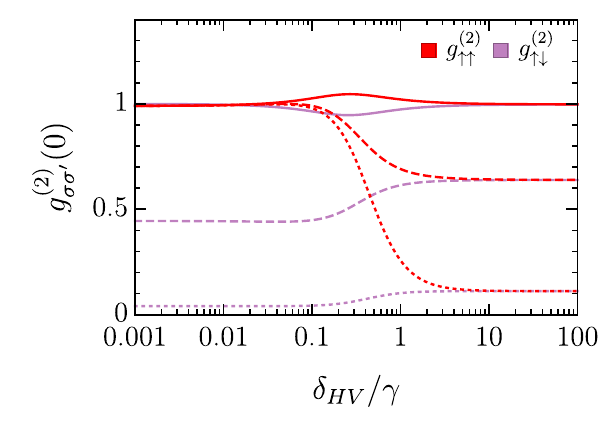}
\caption{ Second-order correlations versus birefringence at resonance. Solid lines: $g_{BL}^2/\gamma\gamma_B=0$,  Dashed-lines: $g_{BL}^2/\gamma\gamma_B=0.25$,  dotted-lines: $g_{BL}^2/\gamma\gamma_B=2$. ($U_1=0.1\gamma$). \new{While for clarity we did not superimpose the curves here, we note that the numerical results from the Lindblad equation with $F=0.01\gamma$ perfectly match the analytical ones [Eq.~\eqref{eq:g2Lin_res}].}} 
\label{fig:g2Feshlindhv} 
\end{figure}

In the opposite limit, when $|\delta_{HV}|\gg\gamma,U_1$, the correlations tend to be equal and one has
\begin{eqnarray}\label{eq:gresonancedhV}
g_{\sigma\sigma,res}^{(2)}&\simeq&g_{\up\dn,res}^{(2)}\simeq\frac{1}{(g_{BL}^2/\gamma\gamma_B+1)^2+U_1^2/4\gamma^2}.
\end{eqnarray}
Here, both dimensionless ratios $g_{BL}^2/\gamma\gamma_B$ and $U_1^2/\gamma^2$ can contribute to the common antibunching.
Figure~\ref{fig:g2Feshlindhv} illustrates the interplay between the limiting cases of Eqs.~\eqref{eq:gresonanceNodhV} and \eqref{eq:gresonancedhV} for several values of the ratio $g_{BL}^2/\gamma\gamma_B$. We can observe that $g_{BL}^2/\gamma\gamma_B=2$ (dotted-lines) already leads to a remarkable antibunching.

\begin{figure}[tbp]
   \includegraphics[width=0.9\linewidth]{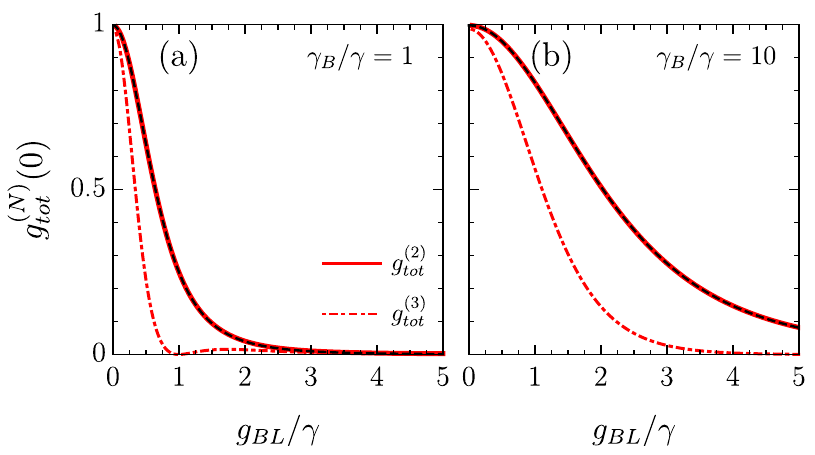}
\caption{ Total intensity correlations at Feshbach resonance versus $g_{BL}$ in presence of large birefringence splitting. Solid and dashed-dotted red lines correspond to the second and third order correlations calculated using Lindblad equation. The dashed black lines correspond to the analytical result of Eq.~\eqref{eq:gresonancedhV}. (a) $\gamma_B=\gamma$, (b) $\gamma_B=10\gamma$. Parameters: $\Delta_{B}=2\Delta_H=0$, $\delta_{HV}=-10\gamma$,  $U_1=0.1\gamma$, $F=0.01\gamma$.} 
\label{fig:gtotFesch} 
\end{figure}

Achieving $g_{\up\dn}^{(2)}\ll1$ and $g_{\sigma\sigma}^{(2)}\ll1$ simultaneously implies that any two-polariton state is blocked.
Thus, Fig.~\ref{fig:g2Feshlindhv} shows that the regime of large birefringence is particularly well suited for the investigation of the Feshbach polariton blockade, 
as proposed in Ref.~\cite{Carusotto_2010}. 
To further illustrate this point, panels (a) and (b) of Fig.~\ref{fig:gtotFesch} display the second and third order total-intensity correlations for such configuration for $\gamma_B/\gamma =1$ and $\gamma_B/\gamma =10$, respectively. 
We see that Eq.~\eqref{eq:gresonancedhV} (dashed black) perfectly matches the results of the numerical calculation of $g_{tot}^{(2)}$ (solid red lines).
The dot-dashed red lines represent the numerically calculated $g_{tot}^{(3)}$. Both correlation functions exhibit an antibunching as $g_{BL}/\gamma$ is increased, which confirms the onset of the polariton blockade effect mediated by the strong nonlinearity associated with the Feshbach resonance.
 

Our analytical expression combined with the coupling $g_{BL}$ introduced in Eq.~\eqref{eq:coupling} can be used to estimate the expected antibunching for given experimental parameters. In particular, the form of $g_{BL}$ suggests that TMD monolayers, in which the biexciton binding energies are large \cite{Mayers2015,Komsa2015,Kidd2016}, are promising candidates for achieving a significant polariton antibunching.
For example, taking $U_1\simeq0$ and using the parameters: $\gamma_B=\gamma=60\mu$eV, $\mathcal{A}=\pi\mu$m$^2$, $X^2=0.5$, $\epsilon_{B,\text{GaAs}}^{XX}=2.94$meV, $\epsilon_{B,\text{MoSe}_2}^{XX}=20$ meV \cite{hao2017neutral}, $m_{X,\text{MoSe}_2}\simeq m_0$, $m_{X,\text{GaAs}}\simeq 0.63m_0$, then the total intensity correlations are $g_{res,\text{GaAs}}^{(2)}\simeq0.70$, and $g_{res,\text{MoSe}_2}^{(2)}\simeq0.29$, according to our result for a large birefringence, Eq.~\eqref{eq:gresonancedhV}.

\subsection{Away from resonance}

In practice, tuning the polariton perfectly on the resonance condition can be challenging, and one might rather have $E_{XX}-2E_L^H
\neq0$.
In order to visualise what are the new signatures expected in an experimental measurement, it is insightful to plot the correlations versus the laser detuning from the driven mode $\Delta_H$.

\begin{figure*}[htbp]  
   \includegraphics[width=\linewidth]{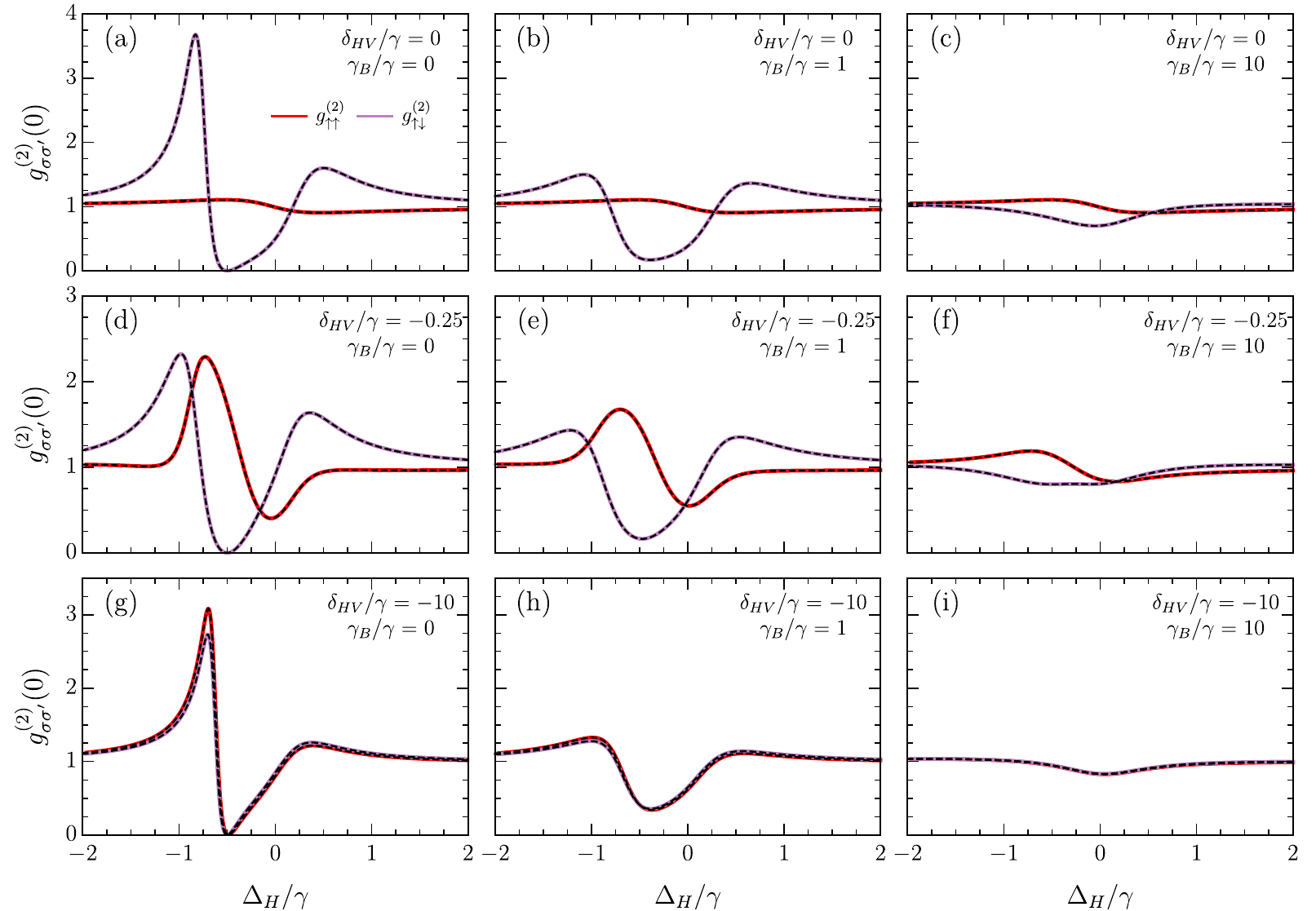}
\caption{\new{Second-order correlations versus $\Delta_H$. First line: no birefringence $\delta_{HV}=0$. (a) $\gamma_B=0$,  (b) $\gamma_B=\gamma$, (c) $\gamma_B=10\gamma$. Second line: small birefringence $\delta_{HV}=-0.25\gamma$. (d) $\gamma_B=0$,  (e) $\gamma_B=\gamma$, (f) $\gamma_B=10\gamma$. Third line: large birefringence   $\delta_{HV}=-10\gamma$. (g) $\gamma_B=0$,  (h) $\gamma_B=\gamma$, (i) $\gamma_B=10\gamma$. The solid colored lines correspond to the numerical results from the Lindblad equation while the dashed-black lines correspond to the analytical results (Eq. \eqref{eq:g+++-Lin_Fesh}).} Other parameters: $E_{XX}-2E_L^H=\gamma$, $g_{BL}=\gamma$, $U_1=0.1\gamma$, $F=0.01\gamma$. }\label{fig:Feshbach}
\end{figure*}

\new{The results corresponding to the case $E_{XX}-2E_L^H=\gamma$ are plotted in Fig.~\ref{fig:Feshbach}.
The purple (red) curves represent $g_{\up\dn}^{(2)}$ ($g_{\sigma\sigma}^{(2)}$) calculated numerically from the Lindblad equation, and the dashed-black lines correspond to the analytical results presented in Eq. \eqref{eq:g+++-Lin_Fesh}.
The different rows correspond to different birefringence splittings. The different columns correspond to different biexciton decay $\gamma_B=0$, $\gamma_B=\gamma$ and $\gamma_B=10\gamma $, respectively.  We can see that when $\gamma_B=0$, $g_{\up\dn}^{(2)}$ always vanishes at $\Delta_H=-\gamma/2$ corresponding to the condition satisfying $\Delta_B=0$.

Figures~\ref{fig:Feshbach}(a,b,c) display the correlations in absence of birefringence. In this case, $g_{\up\up}^{(2)}$ is unaffected by the Feshbach resonance which determines only $g_{\up\dn}^{(2)}$. We can see that $g_{\up\dn}^{(2)}$ exhibits bunching on both side of the resonance. This is an original feature of the present model which does not appear in a model with constant cross-Kerr non-linearity. We note that the right and left bunching peaks would be symmetric if $E_{XX}=2E_{L}^{H}$.  When $\gamma_B$ increases (panels (b) and (c)) the antibunching is reduced and the position of the minimum moves to the right.

Figures~\ref{fig:Feshbach}(d,e,f) show the results for a small birefringence splitting. We can see that $g_{\up\up}^{(2)}$ starts to be affected by the biexciton and exhibits additional bunching/antibunching features with respect to panel (a).
Finally, Figs.~\ref{fig:Feshbach}(g,h,i) correspond to the configuration of a large birefingence splitting. In this case, we see that $g_{\up\up}^{(2)}$ and $g_{\up\dn}^{(2)}$ curves start to overlap, as expected.}

\section{Conclusions}
\label{Sec:Conc}

In conclusion, combining analytics and numerics we have investigated the problem of a weakly driven polariton box. In the presence of birefringence splitting, the system can be modeled as two coupled Kerr resonators with both Kerr and cross-Kerr-like nonlinearities. This allowed us to highlight the possibility of realizing the unconventional blockade with a circularly-polarized drive. In this context, we have obtained the optimal conditions generalizing previous results to the case with nonzero cross-Kerr nonlinearity.

We have also obtained analytical expressions for the second-order correlation functions in the linearly polarized drive configuration, and discussed the limiting cases where the birefringence is large or absent, as well as the regime of weak nonlinearities. 
In particular, the simplified expression of the total-intensity correlation in the latter regime combined with the estimation of $U_1$ from Eq.~\eqref{eq:interactions} allowed us to argue that the recent experiments 
appear to have probed the regime with $U_2\gg U_1$. 

This motivated us to investigate the regime of the polariton Feshbach resonance within an effective two-channel model. In doing so, we have introduced a polariton-biexciton coupling $g_{BL}$ consistent with the cross-Kerr interaction strength in the vicinity of the resonance. At resonance, we demonstrated that the relevant dimensionless parameter characterising the antibunching is $g_{BL}^2/\gamma\gamma_B$.  Combining this resonance with a large birefringence splitting appears promising to achieve strong antibunching in the total intensity correlation functions and reach the blockade regime. 
An experimental challenge in this direction might be to finely tune the polariton mode to the resonance, which would require a precise knowledge of the biexciton energy, a quantity which can be affected by sample-dependent factors such as the thickness of the quantum well or the surroundings of the semiconductor monolayer.

The interplay between polarization and nonlinearity also opens up intriguing perspectives in the high-intensity driving regime such as further exploration of ``dissipative phase transitions'' beyond mean field approximation \cite{Rodrigez2017,Fink2018}.
In a similar context, it would be interesting to see if the cross-Kerr nonlinearities can affect the ``dissipative time crystals'' recently predicted for Kerr resonators dimers \cite{Lledo2019,Seibold2020}.
Finally, for incoherent pumping, the regime of large cross-Kerr nonlinearity $U_2\gg U_1$ combined with arrays of non-birefringent cavities might be of interest for the investigation of bosonic phase separation on a lattice and its potential interplay with photonic spin-orbit-coupling \cite{Sala2015,Nalitov2015}.

\acknowledgments
We acknowledge support from the Australian Research Council Centre of Excellence in Future Low-Energy Electronics Technologies (CE170100039).
JL and MMP are also supported through Australian Research Council Future Fellowships FT160100244 and FT200100619, respectively.

\appendix

\section{Correlation functions}\label{Sec:correlations}
In this appendix, we introduce the different zero-delay correlation functions that we will calculate in the main text.
First, the unnormalized second-order correlation functions for co- and cross-circular polarizations are defined as:
\begin{eqnarray}
G^{(2)}_{\sigma\sigma'}=\langle\hat{L}_\sigma^\dagger\hat{L}_{\sigma'}^\dagger\hat{L}_{\sigma'}\hat{L}_\sigma\rangle,
\end{eqnarray}
and the corresponding normalized second order correlations read
\begin{eqnarray}
g^{(2)}_{\sigma\sigma'}=\frac{G_{\sigma\sigma'}^{(2)}}{\langle\hat{n}_\sigma\rangle\langle\hat{n}_{\sigma'}\rangle},
\end{eqnarray}
where $\hat{n}_\sigma=\hat{L}_\sigma^\dagger\hat{L}_\sigma$. 
We also consider the second-order total intensity (polarization unresolved) correlation function
\begin{equation}\label{eq:g2tot}
g_{tot}^{(2)}=\frac{\sum_{\sigma,\sigma'}G_{\sigma\sigma'}}{\langle\hat{n}_\up+\hat{n}_\dn\rangle^2}=
\frac{\langle(\hat{n}_\up+\hat{n}_\dn)(\hat{n}_\up+\hat{n}_\dn-1)\rangle}{\langle\hat{n}_\up+\hat{n}_\dn\rangle^2}.
\end{equation}

These definitions generalize naturally to the $N$'th order correlation functions. The unnormalized correlation function is
\begin{align}
G^{(N)}_{\sigma_1\cdots \sigma_N}=\langle\hat{L}_{\sigma_1}^\dagger\cdots \hat{L}_{\sigma_N}^\dagger\hat{L}_{\sigma_N}\cdots\hat{L}_{\sigma_1}\rangle,
\end{align}
from which we obtain the normalized correlation function for
co-circularly polarized polaritons
\begin{align}
g^{(N)}_{\sigma\sigma}=\frac{\langle\hat{L}_\sigma^{\dagger N}\hat{L}_\sigma^{N}\rangle}{\langle\hat{n}_\sigma\rangle^{N}}.
\end{align}
and the 
total intensity 
coherence function 
\begin{align}\label{eq:gNtot}
g_{tot}^{(N)}&=&\frac{\sum_{\sigma_1,\dots,\sigma_N}G^{(N)}_{\sigma_1\cdots\sigma_N}} {\langle\hat{n}_\up+\hat{n}_\dn\rangle^N} = \frac{\langle\Pi_{k=0}^{N-1}(\hat{n}_\up+\hat{n}_\dn-k)\rangle}{\langle\hat{n}_\up+\hat{n}_\dn\rangle^N}.
\end{align}

In the wave-function approach the average are calculated as $\langle\hat{O}\rangle=\bra{\psi}\hat{O}\ket{\psi}$. The populations and first order coherences read
\begin{subequations}
\begin{eqnarray}
\bar{n}_\up&=&\sum_{n,m}n|C_{nm}|^2, \\ 
\bar{n}_\dn&=&\sum_{n,m}m|C_{nm}|^2
, \\
\bar{n}_{\up\dn}&=&\sum_{n,m}\sqrt{(n+1)m}C_{n+1,m-1}^*C_{nm}, \\
\bar{n}_{\dn\up}&=&\sum_{n,m}\sqrt{(m+1)n}C_{n-1,m+1}^*C_{nm}.
\end{eqnarray}
\end{subequations}
The second-order coherence functions are of the form:
\begin{subequations}
\begin{eqnarray}
g_{\up\up}^{(2)}&=&\frac{1}{\bar{n}_\up^2}\sum_{n,m}n (n-1)|C_{nm}|^2, \\
g_{\dn\dn}^{(2)}&=&\frac{1}{\bar{n}_\dn^2}\sum_{n,m}m (m-1)|C_{nm}|^2, \\
g_{\up\dn}^{(2)}&=&g_{\dn\up}^{(2)}=\frac{1}{\bar{n}_\up\bar{n}_\dn}\sum_{n,m}nm|C_{nm}|^2, \\
g_{tot}^{(2)}&=&\frac{\sum_{n,m}(n+m) (n+m-1)|C_{nm}|^2}{(\bar{n}_\up+\bar{n}_\dn)^2}.
\end{eqnarray}
\end{subequations}
Upon truncation, these expressions reduce to Eqs.~\eqref{eq:pop} and \eqref{eq:g2} given in the main text.


It is also useful to introduce similar correlations in the linear polarization basis, which can be of practical interest.
The circular and horizontal-vertical polarization annihilation operators are related as follows:
\begin{eqnarray}\label{eq:trans}
    \begin{pmatrix} 
    \hat{L}_H\\
    \hat{L}_V
      \end{pmatrix} &=& \frac{1}{\sqrt{2}}\begin{pmatrix} 
    1 & 1\\-i &i
      \end{pmatrix} \begin{pmatrix} 
    \hat{L}_{\up}\\
    \hat{L}_{\dn}
      \end{pmatrix}.
\end{eqnarray}
For an arbitrary orientation of linear polarization basis, the polariton operators read ($\theta$ is the angle from H-axis, and $\bar{\theta}=\theta+\frac{\pi}{2}$):
\begin{eqnarray}
    \begin{pmatrix} 
    \hat{L}_\theta\\
    \hat{L}_{\bar{\theta}}
      \end{pmatrix} =\frac{1}{\sqrt{2}} \begin{pmatrix} 
    e^{-i\theta} &   e^{i\theta} \\  e^{-i\bar{\theta}} &   e^{i\bar{\theta}}
      \end{pmatrix} \begin{pmatrix} 
    \hat{L}_{\up}\\
    \hat{L}_{\dn}
      \end{pmatrix}.
\end{eqnarray}
The average numbers of linearly polarized polaritons read:
\begin{eqnarray}
\bar{n}_\theta&=&\langle\hat{L}_{\theta}^\dagger\hat{L}_{\theta}\rangle , ~\bar{n}_{\theta=0}= \bar{n}_H, ~\bar{n}_{\theta=\frac{\pi}{2}}= \bar{n}_V.
\end{eqnarray}
and one has
\begin{eqnarray}
\bar{n}_{\theta}&=&\frac{1}{2}\left(\bar{n}_\up+\bar{n}_\dn+e^{2i\theta}\bar{n}_{\up\dn}+\bar{n}_{\dn\up}e^{-2i\theta}\right).
\end{eqnarray}

The $N$-th order normalized coherence functions for linearly polarized polaritons are:
\begin{eqnarray}
g^{(N)}_{HH}=\frac{G_{HH}^{(N)}}{\langle\hat{L}_H^\dagger\hat{L}_H\rangle^N}, ~~ g^{(N)}_{VV}=\frac{G_{VV}^{(N)}}{\langle\hat{L}_V^\dagger\hat{L}_V\rangle^N},
\end{eqnarray}
with
\begin{eqnarray}
G_{HH}^{(N)}=\langle\hat{L}_H^{\dagger N}\hat{L}_H^N\rangle, ~~ G_{VV}^{(N)}=\langle\hat{L}_V^{\dagger N}\hat{L}_V^N\rangle,
\end{eqnarray}
where operator products $\hat{L}_\theta^{\dagger N}\hat{L}_\theta^N$ can be expressed in terms of the operators $\hat{L}_{\up,\dn}^{\dagger}$ and $\hat{L}_{\up,\dn}$ as
\begin{equation}\nonumber
\hat{L}_\theta^{\dagger N}\hat{L}_\theta^N =\sum_{k,l=0}^N  \binom{N}{k} \binom{N}{l} \frac{e^{2i\theta (l-k)}}{2^N} \hat{L}_{\up}^{\dagger N-k}\hat{L}_{\dn}^{\dagger k}\hat{L}_{\up}^{N-l}\hat{L}_{\dn}^{l}.
\end{equation}
Finally, the normalized second-order correlation for cross-linear polarization $g_{HV}^{(2)}$ is defined as:
\begin{equation}
g_{HV}^{(2)}=\frac{\langle\hat{L}_H^\dagger\hat{L}_{V}^\dagger\hat{L}_{V}\hat{L}_H\rangle}{\langle\hat{L}_H^\dagger\hat{L}_H\rangle\langle\hat{L}_V^\dagger\hat{L}_V\rangle}.
\end{equation}

\onecolumngrid
\section{Linearly polarized drive: effective wave function versus Lindblad equation}\label{Appendix:Lin}
In this Appendix, we discuss a subtle difference between the effective wave function and Lindblad formalism for the H-polarized drive configuration. We highlight that while the symmetry of the Hamiltonian prohibits the occupation of odd V Fock states in the effective wave-function approach, this strict prohibition disappears  in the Lindblad formalism. We then use a pertubative expansion of the density matrix to obtain accurate analytical results for the observable quantities $\bar{n}_V$ and $g^{(2)}_{HV}$.

When the drive is H polarized, $F_\up=F_\dn=F$, the use of \eqref{eq:trans} allows us to rewrite the Hamiltonian \eqref{eq:HtotKerr} in the H-V basis as
\begin{eqnarray}  
   \hat{H}&=&\left(\Delta+\delta_{HV}\right)\nonumber \hat{L}_{H}^\dagger\hat{L}_{H}+\left(\Delta-\delta_{HV}\right) \hat{L}_{V}^\dagger\hat{L}_{V}
   +F_{H}\left(\hat{L}_{H}
   +\hat{L}_{H}^{\dagger}\right) + U_1 \left[\hat{L}_{H}^\dagger\hat{L}_{V}^\dagger\hat{L}_{V}\hat{L}_{H} \right]\\ 
   &&+\frac{U_{1}+U_2}{4} \left[\hat{L}_{H}^\dagger\hat{L}_{H}^\dagger\hat{L}_{H}\hat{L}_{H} +\hat{L}_{V}^\dagger\hat{L}_{V}^\dagger\hat{L}_{V}\hat{L}_{V}\right] 
  -\frac{U_{1}-U_2}{4} \left[\hat{L}_{H}^\dagger\hat{L}_{H}^\dagger\hat{L}_{V}\hat{L}_{V} +\hat{L}_{V}^\dagger\hat{L}_{V}^\dagger\hat{L}_{H}\hat{L}_{H}\right], \label{eq:HamLinear}
\end{eqnarray}
with $F_H=\sqrt{2}F$. Here, we see that when $U_1=U_2$, the {H-V} subspaces decouple and the Fock states $\ket{n,m}_{HV}$ with $m\neq0$ remain unoccupied since $F_V=0$.
When $U_1\neq U_2$ the two subspaces are coupled by the last term of \eqref{eq:HamLinear}. Importantly, this coupling is two-body as it remove two excitations in one subspace to create two excitations in the other. As a consequence, in our effective wave-function approach (which neglects quantum jumps), the states $\ket{n,m}_{HV}$ with odd $m$ cannot be occupied, and thus when truncated to $n+m\leq2$, $\bar{n}_V$ is fully determined by the occupation of $\ket{0,2}_{HV}$.

However, this 
is no longer true in the Lindblad formalism where the dissipation induces additional couplings.
This can be understood by looking at the evolution equations of the diagonal elements of the density matrix in the HV Fock basis $\rho_{n,m,n,m}^{HV}=\bra{nm}\hat{\rho}\ket{nm}_{HV}$.
First, we note that since we assumed a polarization independent decay, the dissipation term in the Lindblad equation is diagonal in either basis and one has
\begin{align}
  \gamma \sum_{\sigma=\up,\dn} \left(   \hat{L}_{\sigma}\hat{\rho}\hat{L}_{\sigma}^{\dagger} - \frac{1}{2}\{\hat{L}_{\sigma}^{\dagger}\hat{L}_{\sigma},\hat{\rho}\} \right)=  \gamma \sum_{\lambda={\text{H,V}}} \left(   \hat{L}_{\lambda}\hat{\rho}\hat{L}_{\lambda}^{\dagger} - \frac{1}{2}\{\hat{L}_{\lambda}^{\dagger}\hat{L}_{\lambda},\hat{\rho}\} \right) .
\end{align}
Then, using Eq.~\eqref{eq:Lindblad}, the evolution of the diagonal elements in the HV basis read
\begin{align}\label{eq:LindbladHV}
  \hbar \dot{\rho}_{n,m,n,m}^{HV}  &=-i \bra{n,m} \left[ \hat{H},\hat{\rho}\right]\ket{n,m}_{HV}+\gamma (n+1) \rho_{n+1,m,n+1,m}^{HV} +\gamma (m+1) \rho_{n,m+1,n,m+1}^{HV}-\gamma ( n +m) \rho_{n,m,n,m}^{HV} .
\end{align}
The first term encodes the Hamiltonian evolution, and thus cannot induce transitions to $\rho_{n,m,n,m}^{HV}$ with odd $m$ as explained above. The other terms are related to the coupling with the environment as described in the Lindblad equation.  The last term is related to the decay to the outside, a phenomenon encoded in the effective wave-function approach. The second and third terms encode transitions between levels and originate from the $\hat{L}_{\sigma}\hat{\rho}\hat{L}_{\sigma}^{\dagger} $ Lindblad terms. The third term can induce population of the levels with odd $m$ index. This can be interpreted as an explicit symmetry breaking due to the environment.

\subsection{Perturbative expansion of the density matrix}\label{App:pert}

Here, we perform a perturbative expansion of the steady state density matrix to calculate correlations in HV basis in the low drive limit.
 First, we recall that in terms of the density matrix elements in the HV basis, one has the relations:
  \begin{subequations}
  \begin{align} 
  \bar{n}_H&= \sum_{n,m} n\rho_{n,m,n,m}^{HV},\\
  \bar{n}_V&= \sum_{n,m} m \rho_{n,m,n,m}^{HV},\label{eq:nbarVappB}\\
g_{VV}^{(2)}&= \frac{1}{\bar{n}_V^2}\sum_{n,m} m(m-1)\rho_{n,m,n,m}^{HV},\\
g_{HH}^{(2)}&= \frac{1}{\bar{n}_H^2}\sum_{n,m} n(n-1)\rho_{n,m,n,m}^{HV}, \label{eq:ghhappB}
\\
g_{HV}^{(2)}&= \frac{1}{\bar{n}_H\bar{n}_V}\sum_{n,m} nm\rho_{n,m,n,m}^{HV}.
 \end{align}
  \end{subequations}
To estimate Eqs. \eqref{eq:nbarVappB}-\eqref{eq:ghhappB} one has to evaluate density matrix elements up to the fourth order in $F$.
 Assuming $ \rho_{0,0,0,0}^{HV}\simeq 1$, one has to first order in $F$:
 \begin{eqnarray} 
  \rho_{0,0,1,0}^{HV}= \rho_{1,0,0,0}^{HV*}\simeq i\frac{ F_H}{\gamma/2-i(\Delta+\delta_{HV})} \rho_{0,0,0,0}^{HV},
\end{eqnarray}
to second order:
 \begin{subequations}\label{eq:F2}
 \begin{align} 
\rho_{0,0,2,0}^{HV} &= \rho_{0,0,2,0}^{HV*}\simeq \left( i\frac{ U_2-U_1}{2} \rho_{0,0,0,2}^{HV}+i F_H\sqrt{2}  \rho_{0,0,1,0}^{HV} \right) \left[\gamma-i\frac{ U_1+U_2}{2} -i2(\Delta+\delta_{HV})\right]^{-1} , 
\\
 \rho_{0,0,0,2}^{HV}&= \rho_{0,2,0,0}^{HV*} \simeq  i\frac{ U_2-U_1}{2} \rho_{0,0,2,0}^{HV}\left[\gamma-i\frac{ U_1+U_2}{2}-i2(\Delta-\delta_{HV})\right]^{-1} ,
 \\
   \rho_{1,0,1,0}^{HV}&= \rho_{1,0,1,0}^{HV*}\simeq i\frac{ F_H}{\gamma} ( \rho_{0,0,1,0}^{HV} - \rho_{1,0,0,0}^{HV}),
\end{align}
\end{subequations}
to third order:
\begin{subequations}\label{eq:F3}
 \begin{align} 
 \rho_{1,0,2,0}^{HV}&= \rho_{2,0,1,0}^{HV*}\simeq \left( i F_H\left(  \rho_{1,0,1,0}^{HV} \sqrt{2}-  \rho_{0,0,2,0}^{HV}\right) +i \frac{U_2-U_1}{2}  \rho_{1,0,0,2}^{HV}\right) \left[\frac{3}{2}\gamma-i\frac{ U_1+U_2}{2}-i(\Delta+\delta_{HV})\right]^{-1} ,
 \\
 \rho_{1,0,0,2}^{HV}&= \rho_{0,2,1,0}^{HV*}\simeq \left(- i F_H \rho_{0,0,0,2}^{HV} +i \frac{U_2-U_1}{2}  \rho_{1,0,2,0}^{HV}\right)  \left[\frac{3}{2}\gamma-i\frac{ U_1+U_2}{2}-i(3\delta_{HV}-\Delta)\right]^{-1} ,
\end{align}
\end{subequations}
to fourth order:
\begin{subequations}\label{eq:F4}
 \begin{align} 
  \rho_{0,1,0,1}^{HV}&\simeq 2\rho_{0,2,0,2} \label{eq:rho0101}\\
 \rho_{0,2,0,2}^{HV}&\simeq i\frac{U_2-U_1}{4\gamma}( \rho_{0,2,2,0}^{HV}-\rho_{2,0,0,2}^{HV}),\\
  \rho_{2,0,2,0}^{HV}&\simeq \left(i F_H \sqrt{2} (\rho_{2,0,1,0}^{HV}-\rho_{1,0,2,0}^{HV})+i\frac{U_2-U_1}{2}( \rho_{2,0,0,2}^{HV}-\rho_{0,2,2,0}^{HV})\right)(2\gamma)^{-1}, \\
    \rho_{2,0,0,2}^{HV}&= \rho_{0,2,2,0}^{HV*}\simeq\left(-i F_H \sqrt{2} \rho_{1,0,0,2}^{HV}+i\frac{U_2-U_1}{2}( \rho_{2,0,2,0}^{HV}-\rho_{0,2,0,2}^{HV})\right)\left[2\gamma+4i\delta_{HV}\right]^{-1}.
\end{align}
\end{subequations}
Equation~\eqref{eq:rho0101}, which is related to the third term in \eqref{eq:LindbladHV}, implies that the probability to occupy the 1-V polariton state is twice  the one to occupy the 2-V state. Solving Eqs.~\eqref{eq:F2}-\eqref{eq:F4}, one obtains
 \begin{eqnarray} 
  \rho_{0,1,0,1}^{HV}=  2\rho_{0,2,0,2}^{HV} \simeq\frac{16 F^4}{ 4(\Delta+\delta_{HV})^2+\gamma^2} \frac{ \left(U_1-U_2\right)^2}{ \left(D_1 D_2-4\delta_{HV}^2\right)^2+\gamma^2\left( D_1^2+D_2^2+8\delta_{HV}^2\right)+\gamma^4 }=\bar{n}_{V,wf},
\end{eqnarray}
and thus, using Eq.~\eqref{eq:nbarVappB} and the fact that $\rho^{HV}_{1,1,1,1}=0$ at this order, we find that the V population is exactly twice the result obtained in the wavefunction approach \eqref{eq:ny}  $\bar{n}_V\simeq 2 \bar{n}_{V,wf}$. One can then deduce the second order correlation $g_{VV}^{(2)}$
 \begin{eqnarray} 
g_{VV}^{(2)} \simeq  \frac{2\rho_{0,2,0,2}^{HV}}{\left(\rho_{0,1,0,1}^{HV}+2\rho_{0,2,0,2}^{HV}\right)^2}=  \frac{1}{2\bar{n}_V}.
\end{eqnarray}
This result does not match the one of the effective wavefunction approach $g_{VV,wf}^{(2)}=1/\bar{n}_{V,wf}$. Nonetheless, we can see that both results diverge as $1/F^4$ in the low drive limit.

One can also calculate the second-order correlation in the same polarization as the drive $g_{HH}^{(2)}$, which is finite in the low drive limit and reads
\begin{align} \label{eq:gHHLin}
g_{HH}^{(2)}  \simeq  \frac{2\rho_{2,0,2,0}^{HV}}{(\rho_{1,0,1,0}^{HV})^2}=\frac{\left((\Delta+\delta_{HV})^2+\gamma ^2/4\right) \left(\left(4(\Delta-\delta_{HV})+U_1+U_2)\right)^2+4\gamma^2\right)}{ \left(D_1 D_2-4\delta_{HV}^2\right)^2+\gamma^2\left( D_1^2+D_2^2+8\delta_{HV}^2\right)+\gamma^4 }.
\end{align}
Here, the right hand side of Eq. \eqref{eq:gHHLin} can be equivalently calculated in both approaches since quantum jumps do not participate in the calculation of the relevant density matrix elements to the leading order. We also note that the ratio $g_{HH}^{(2)}/g_{tot}^{(2)}$ can be expressed as
\begin{align} \label{eq:gHHLinb}
\frac{g_{HH}^{(2)}}{g_{tot}^{(2)}}\simeq 1-\frac{(U_1-U_2)^2/2}{(2 \Delta_V +U_1)^2+(2 \Delta_V+ U_2)^2+2\gamma^2},
\end{align}
where we have introduced $\Delta_V=\Delta-\delta_{HV}$. Equation~\eqref{eq:gHHLinb} highlights that $g_{HH}^{(2)}\leq g_{tot}^{(2)}$ and that $g_{HH}^{(2)}/ g_{tot}^{(2)}<1$ is a signature of unequal interactions $U_1\neq U_2$.\\

\begin{figure}[tbp]
   \includegraphics[scale=1]{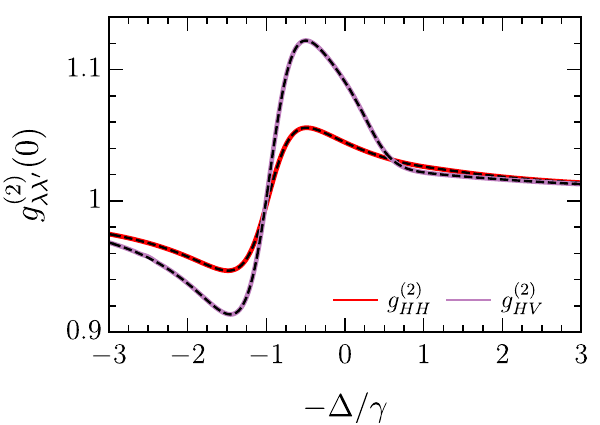}
\caption{Correlation functions in the linear polarization versus detuning $\Delta$ under weak H drive. Same parameters as in Fig \ref{fig:g2lin} of the main text ($U_1=0.01\gamma$, $U_2=0.1\gamma$, $\delta_{HV}=-\gamma$, $F=0.01\gamma$).
The solid colored lines and dot-dashed red lines correspond to the numerical results while the dashed-black lines correspond to the analytical results in Eqs.~\eqref{eq:gHHLin} and \eqref{eq:ghvLind}. }
\label{fig:gHVlin} 
\end{figure}

The analytical estimation of $g_{HV}^{(2)} $ is slightly more involved as it requires to go to higher order in the expansion. To the leading order it reads
 \begin{align} \nonumber
 g_{HV}^{(2)}\simeq& \frac{\rho_{1,1,1,1}^{HV}+2\rho_{1,2,1,2}^{HV}}{\rho_{1,0,1,0}^{HV}\left(\rho_{0,1,0,1}^{HV}+2\rho_{0,2,0,2}^{HV}\right)}.
\end{align}
To estimate it, one must go beyond the truncation to $n+m=2$, and one has to solve an additional set of equations.
Elements of third order $ F^3$:
\begin{subequations}\label{eq:F3b}
 \begin{align} 
 \rho_{0,0,3,0}^{HV}&= \rho_{3,0,0,0}^{HV*}\simeq \left(i \frac{\sqrt{3}}{2}(U_2-U_1) \rho_{0,0,1,2}^{HV} +i\sqrt{3}F_H \rho_{0,0,2,0}^{HV}\right) \left[\frac{3}{2}\gamma-i\frac{3}{2}(U_1+U_2)-i 3\Delta_H\right]^{-1},
 \\
 \rho_{0,0,1,2}^{HV}&= \rho_{1,2,0,0}^{HV*}\simeq \left(iF_H \rho_{0,0,0,2}^{HV}+i \frac{\sqrt{3}}{2}(U_2-U_1) \rho_{0,0,3,0}^{HV} \right)\left[\frac{3}{2}\gamma-2i U_1-i\frac{1}{2}(U_1+U_2)-i \Delta_H-2i \Delta_V\right]^{-1},
\end{align}
\end{subequations}
 Elements of order $ F^4$:
\begin{subequations}\label{eq:F4b}
\begin{align}
 \rho_{1,0,3,0}^{HV}&= \rho_{3,0,1,0}^{HV*}\simeq \left(-iF_H  \rho_{0,0,3,0}^{HV} + i \sqrt{3} F_H  \rho_{1,0,2,0}^{HV}+i\frac{\sqrt{3}}{2}(U_2-U_1) \rho_{1,0,1,2}^{HV}\right)\left[2\gamma-i\frac{3}{2}(U_1+U_2)-i 2 \Delta_H\right]^{-1},
 \\
 \rho_{1,0,1,2}^{HV}&= \rho_{1,2,1,0}^{HV*}\simeq \left(-iF_H  (\rho_{0,0,1,2}^{HV} - \rho_{1,0,0,2}^{HV} ) +i\frac{\sqrt{3}}{2}(U_2-U_1) \rho_{1,0,3,0}^{HV}\right)\left[2\gamma-i 2 U_1-i\frac{3}{2}(U_1+U_2)-i2 \Delta_V\right]^{-1},
\end{align}
\end{subequations}
Elements of fifth order $ F^5$:
\begin{subequations}\label{eq:F5}
 \begin{align} 
 \rho_{2,0,3,0}^{HV}&= \rho_{3,0,2,0}^{HV*}\simeq \left( i F_H(\sqrt{3}  \rho_{2,0,2,0}^{HV}- \sqrt{2}  \rho_{1,0,3,0}^{HV}) +i \frac{U_2-U_1}{2}(\sqrt{3} \rho_{2,0,1,2}^{HV}- \rho_{0,2,3,0}^{HV} ) \right)\left[\frac{5}{2}\gamma-i(U_1+U_2)-i \Delta_H\right]^{-1}, \\ 
 \rho_{2,0,1,2}^{HV}&= \rho_{1,2,2,0}^{HV*}\simeq\left( i F_H(\sqrt{2}  \rho_{1,0,1,2}^{HV}-  \rho_{2,0,0,2}^{HV}) +i \frac{U_2-U_1}{2}(\sqrt{3} \rho_{2,0,3,0}^{HV}- \rho_{0,2,1,2}^{HV} ) \right) \left[\frac{5}{2}\gamma-i2 U_1+i \Delta_H-i2 \Delta_V\right]^{-1} ,
 \\
  \rho_{0,2,3,0}^{HV}&= \rho_{3,0,0,2}^{HV*}\simeq \left( i F_H \sqrt{3}  \rho_{0,2,2,0}^{HV} +i \frac{U_2-U_1}{2}(\sqrt{3} \rho_{0,2,1,2}^{HV}- \rho_{2,0,3,0}^{HV} ) \right)\left[\frac{5}{2}\gamma-i(U_1+U_2)-i3 \Delta_H+i2 \Delta_V\right]^{-1} ,
  \\ 
 \rho_{0,2,1,2}^{HV}&= \rho_{1,2,0,2}^{HV*}\simeq \left( i F_H   \rho_{0,2,0,2}^{HV} +i \frac{U_2-U_1}{2}(\sqrt{3} \rho_{0,2,3,0}^{HV}- \rho_{2,0,1,2}^{HV} ) \right) \left[\frac{5}{2}\gamma-i2U_1-i \Delta_H\right]^{-1} ,
 \\
    \rho_{1,1,0,1}^{HV}&=  \rho_{0,1,1,1}^{HV*}\simeq \left(-i  F_H  \rho_{0,1,0,1}^{HV} + 2 \gamma  \rho_{0,2,1,2}^{HV}\right) \left[\frac{3}{2}\gamma+iU_1+i(\Delta+\delta_{HV})\right]^{-1},
\end{align}
\end{subequations}
Elements of sixth order $ F^6$:
\begin{subequations}\label{eq:F6}
 \begin{align} 
  \rho_{1,2,3,0}^{HV}&=\rho_{3,0,1,2}^{HV*}\simeq \left( i F_H(\sqrt{3}  \rho_{1,2,2,0}^{HV}-  \rho_{0,2,3,0}^{HV}) +i \frac{U_2-U_1}{2}\sqrt{3}( \rho_{1,2,1,2}^{HV}- \rho_{3,0,3,0}^{HV} ) \right)\left[3\gamma+i(U_1-U_2)-i4\delta_{HV}\right]^{-1},
  \\
  \rho_{3,0,3,0}^{HV}&\simeq\left( i F_H\sqrt{3} ( \rho_{3,0,2,0}^{HV}-  \rho_{2,0,3,0}^{HV}) +i \frac{U_2-U_1}{2}\sqrt{3}( \rho_{3,0,1,2}^{HV}- \rho_{1,2,3,0}^{HV} ) \right) \left[3\gamma\right]^{-1},
  \\ 
  \rho_{1,2,1,2}^{HV}&\simeq\left( i F_H( \rho_{1,2,0,2}^{HV}-  \rho_{0,2,1,2}^{HV}) +i \frac{U_2-U_1}{2}\sqrt{3}( \rho_{1,2,3,0}^{HV}- \rho_{3,0,1,2}^{HV} ) \right) \left[3\gamma\right]^{-1},
  \\
  \rho_{1,1,1,1}^{HV} &\simeq i \frac{F_H}{2\gamma} ( \rho_{1,1,0,1}^{HV}-\rho_{0,1,1,1}^{HV})+  \rho_{1,2,1,2}^{HV}.
\end{align}
\end{subequations}
Solving \eqref{eq:F3b}-\eqref{eq:F6}, one obtains 
 \begin{align} \nonumber
 g_{HV}^{(2)}=& \frac{\left(\gamma ^2+4 \Delta_{H}^2\right)}{| -3 i \gamma +2 \Delta_{H}+2 U_1 |^2}\\
 & \times \frac{A}{|4 \Delta_{H}^2+4 \Delta_{H} (-2 i \gamma +2 \Delta_V+3 U_1+U_2)+4 \Delta_V (-i \gamma +U_1+U_2)+(2 U_1-i \gamma ) (-3 i \gamma +2
   U_1+4 U_2)|^2}, \label{eq:ghvLind}
\end{align}
with
 \begin{align} \label{eq:A}
A=&81 \gamma ^4+16 \Delta_H^4+64 \Delta_H^3 \Delta_V+80 \Delta_H^3 U_1+32 \Delta_H^3 U_2+72 \gamma ^2 \Delta_H^2+64 \Delta_H^2 \Delta_V^2+208 \Delta_H^2 \Delta_V U_1+80 \Delta_H^2 \nonumber
   \Delta_V U_2
   \\
   &+140 \Delta_H^2 U_1^2+128 \Delta_H^2 U_1 U_2+16 \Delta_H^2 U_2^2+96 \Delta_H \Delta_V^2 U_1+32 \Delta_H \Delta_V^2 U_2+144 \gamma ^2 \Delta_H
   \Delta_V+168 \Delta_H \Delta_V U_1^2\nonumber
   \\
   &+200 \Delta_H \Delta_V U_1 U_2+16 \Delta_H \Delta_V U_2^2+80 \Delta_H U_1^3+168 \Delta_H U_1^2 U_2+108 \gamma ^2
   \Delta_H U_1+48 \Delta_H U_1 U_2^2+96 \gamma ^2 \Delta_H U_2\nonumber
   \\
   &+144 \gamma ^2 \Delta_V^2+44 \Delta_V^2 U_1^2+8 \Delta_V^2 U_1 U_2+12 \Delta_V^2 U_2^2+48
   \Delta_V U_1^3+72 \Delta_V U_1^2 U_2+180 \gamma ^2 \Delta_V U_1+40 \Delta_V U_1 U_2^2\nonumber
   \\
   &+132 \gamma ^2 \Delta_V U_2+16 U_1^4+48 U_1^3 U_2+72 \gamma ^2 U_1^2+44 U_1^2 U_2^2+60 \gamma ^2 U_1 U_2+51 \gamma ^2 U_2^2.
\end{align}
To illustrate the validity of the analytical results, in Figure \ref{fig:gHVlin}, we have plotted $g_{HH}^{(2)}$ and $g_{HV}^{(2)}$ using the same parameters as in figure \ref{fig:g2lin}.

We conclude this appendix by emphasising that the failure of the wave-function approach here, is related to the asymmetric drive with respect to $\{$H,V$\}$ subspaces \textit{and} the two-body nature of the {HV} coupling. In the case of a circularly polarized drive investigated in Sec.~\ref{Sec:Circ}, the drive is asymmetric with respect to the $\{\up,\dn\}$ subspaces but the coupling is one-body.

\section{Two-channel model of the Feshbach resonance}\label{Sec:2chann}

\subsection{Lindblad equation in the two-channel model}
\label{Sec:Lindblad_2chann}

In the two-channel model, the open-dissipative system is described by the following Lindblad equation:
\begin{align}\nonumber
  \hbar \frac{\partial\hat{\rho}}{\partial t}  &=-i \left[ \hat{H},\hat{\rho}\right]+\gamma \sum_{\sigma} \left(   \hat{L}_{\sigma}\hat{\rho}\hat{L}_{\sigma}^{\dagger} - \frac{1}{2}\{\hat{L}_{\sigma}^{\dagger}\hat{L}_{\sigma},\hat{\rho}\} \right)+\gamma_B  \left(   \hat{B}\hat{\rho}\hat{B}^{\dagger} - \frac{1}{2}\{\hat{B}^{\dagger}\hat{B},\hat{\rho}\} \right) , \label{eq:Lindblad2}
\end{align}
where the Hamiltonian $\hat{H}$ is given in Eq.~\eqref{eq:Ham_2chann} of the main text. $\gamma$ is the polariton decay out of the cavity, and $\gamma_B$ is the biexciton nonradiative decay. The corresponding steady state density matrix can be expanded in the Fock basis as
\begin{eqnarray}
\hat{\rho}_{ss} &=&   \sum_{\substack{n,m,o,\\n',m',o'}} \rho_{\substack{n, m, o,n',m',o'}} \ket{n,m,o}\bra{n' ,m',o'} ,
\end{eqnarray}
where the indices $n,m,o$ are associated with the $\sigma=\{\up,\dn\}$ polariton and biexciton subspaces.

\subsection{Wave-function approach in the two-channel model}\label{Sec:WF_2chann}

Similarly to what we did in Section \ref{subsec:WF}, we introduce the effective non Hermitian Hamiltonian:

\begin{eqnarray}
\hat{H}_{\rm eff}=\hat{H}-i\frac{\gamma}{2} \sum_{\sigma} \hat{L}_{\sigma}^{\dagger}\hat{L}_{\sigma}-i\frac{\gamma_B}{2} \hat{B}^{\dagger}\hat{B}.
\end{eqnarray}
Here, the wave function can be expanded as $ \ket{\psi}=\sum_{n,m,o}C_{nmo}\ket{n,m,o}$.
As previously, analytic expressions can be obtained in the low driving regime by using a truncated ansatz with $n+m+2o\leq2$. 
The resulting $C_{nmo}$ evolution equations read:
\begin{subequations}\label{eq:evol2}
\begin{eqnarray}
i\dot{C}_{100}&=&C_{000} F_{\up}^*+C_{010} \delta_{HV}+C_{100} \tilde{\Delta}+C_{110} F_{\dn}+\sqrt{2} C_{200} F_{\up}, \\
i\dot{C}_{010}&=&C_{000} F_{\dn}^*+C_{010} \tilde{\Delta}+\sqrt{2} C_{002} F_{\dn}+C_{100} \delta_{HV}+C_{110} F_{\up}, \\
i\dot{C}_{200}&=&\sqrt{2} C_{100} F_{\up}^*+\sqrt{2} C_{110} \delta_{HV}+C_{200} (2 \tilde{\Delta}+U_1), \\
i\dot{C}_{020}&=&\sqrt{2} C_{010} F_{\dn}^*+C_{020} (2 \tilde{\Delta}+U_1)+\sqrt{2} C_{110} \delta_{HV}, \\
i\dot{C}_{110}&=&C_{001} g_{BL}+C_{010} F_{\up}^*+\sqrt{2} \delta_{HV} (C_{002}+C_{200})+C_{100} F_{\dn}^*+2 C_{110} \tilde{\Delta}, \\
i\dot{C}_{001}&=&C_{001} \tilde{\Delta}_B+C_{110} g_{BL},
\end{eqnarray}
\end{subequations}
with $\tilde{\Delta}_B=\Delta_B-i\gamma_B/2$.
As in Sec.~\ref{subsec:WF}, we impose $C_{000}=1$ and calculate the steady state solution of Eq.~\eqref{eq:evol2}.
The polariton populations and correlations are then obtained by making use of Eqs.~\eqref{eq:pop} and \eqref{eq:g2} with the replacement $C_{nm}\rightarrow C_{nm0}$.

\subsection{Second-order correlation functions}\label{Sec:g2_2chann}

Here we provide the analytical formula for the second order correlation in the two-channel model for a linearly polarized drive. In the low drive limit, one obtains

\begin{subequations}\label{eq:g+++-Lin_Fesh}
\begin{eqnarray}  \label{eq:g++Lin_Fesh}
g_{\sigma\sigma}^{(2)}&=&\frac{ \gamma ^2+4 (\Delta +\delta_{HV})^2}{D} \left(\gamma_B^2 \left(\gamma ^2+4 (\Delta-\delta_{HV} )^2\right)+4\Delta_B^2(\gamma ^2+(2 \Delta -2 \delta_{HV}-\frac{g_{BL}^2}{\Delta_B})^2)+4 \gamma  \gamma_B g_{BL}^2\right), \\ \label{eq:g+-Lin_Fesh}
g_{\up\dn}^{(2)}&=&\frac{ \gamma ^2+4 (\Delta +\delta_{HV})^2}{D} \left(\gamma_B^2+4 \Delta_B^2\right) \left(\gamma ^2+(2 \Delta -2 \delta_{HV}+U_1)^2\right),\\ \nonumber
   D&=&4 \Delta_B^2 \left(8 \delta_{HV}^2 \left(\gamma ^2-2 \Delta  (2 \Delta +U_1)\right)+16 \delta_{HV}^4+\left(\gamma ^2+4 \Delta ^2\right) \left(\gamma ^2+(2 \Delta
   +U_1)^2\right)\right)\\  \nonumber
   && +16 \Delta_B g_{BL}^2 \left(2 \delta_{HV}^2 (2 \Delta +U_1)-\Delta  \left(\gamma ^2+(2 \Delta +U_1)^2\right)\right)\\ \nonumber
   && +4 g_{BL}^4 \left(\gamma ^2+(2 \Delta
   +U_1)^2\right)\\ \nonumber
   && +4 \gamma 
   \gamma_B g_{BL}^2 \left(\gamma ^2+4 \delta_{HV}^2+(2 \Delta +U_1)^2\right)\\\nonumber
   && +\gamma_B^2 \left(8 \delta_{HV}^2 \left(\gamma ^2-2 \Delta  (2 \Delta
   +U_1)\right)+16 \delta_{HV}^4+\left(\gamma ^2+4 \Delta ^2\right) \left(\gamma ^2+(2 \Delta +U_1)^2\right)\right).
\end{eqnarray}
\end{subequations}

In the limit of zero nonradiative biexciton decay, the expressions \eqref{eq:g+++-Lin_Fesh} reduce to:
\begin{subequations}\label{eq:g+++-LinNogammaB}
\begin{align}
g_{\sigma\sigma,\gamma_B=0}^{(2)}&=\frac{\left(4(\Delta+\delta_{HV})^2+\gamma ^2\right) \left((2 (\Delta-\delta_{HV})-\frac{g_{BL}^2}{\Delta_B})^2+\gamma ^2\right)}{ \left((2 \Delta
   +U_1)  (2 \Delta -\frac{g_{BL}^2}{\Delta_B})-4\delta_{HV}^2\right)^2+\gamma^2\left( (2 \Delta
   +U_1)^2+ (2 \Delta -\frac{g_{BL}^2}{\Delta_B})^2+8\delta_{HV}^2\right)+\gamma^4 },\\ 
g_{\up\dn,\gamma_B=0}^{(2)}&=\frac{\left(4 (\Delta+\delta_{HV})^2+\gamma ^2\right) \left((2(\Delta-\delta_{HV})+U_1)^2+\gamma ^2\right)}{ \left((2 \Delta
   +U_1)  (2 \Delta -\frac{g_{BL}^2}{\Delta_B})-4\delta_{HV}^2\right)^2+\gamma^2\left( (2 \Delta
   +U_1)^2+ (2 \Delta -\frac{g_{BL}^2}{\Delta_B})^2+8\delta_{HV}^2\right)+\gamma^4 }.
\end{align}
\end{subequations}
One can observe a one-to-one correspondence between Eq.~\eqref{eq:g+++-LinNogammaB} and Eq.~\eqref{eq:g+++-Lin} upon the replacement $U_2\leftrightarrow -\frac{g_{BL}^2}{\Delta_B}$. This is consistent with the behaviour of $\alpha_2$ given in Eq.~\eqref{eq:alpha2} in the vicinity of its pole. $g_{\up\dn,\gamma_B=0}^{(2)}$ vanishes when $\Delta_B \rightarrow0$ as it should.

For nonzero biexciton decay, it is insightful to consider the limiting cases of no and large birefringence splitting as we did in Sec. \ref{Sec:Lin} for the Kerr model. 
In absence of birefringence \eqref{eq:g++Lin_Fesh} reduces to Eq.~\eqref{eq:g++nodHV}, and \eqref{eq:g+-Lin_Fesh} becomes
\begin{eqnarray} \label{eq.C3}
g_{\up\dn}^{(2)}&=&\frac{\left(\gamma_B^2+4 \Delta_{B}^2\right)\left(\gamma ^2+4 \Delta ^2\right) }{\gamma_B^2 \left(\gamma ^2+4 \Delta^2\right)+4 \left(\left(g_{BL}^2-2 \Delta \Delta_B\right)^2+\gamma ^2 \Delta_B^2+\gamma  \gamma_B g_{BL}^2\right) },
\end{eqnarray}
while in the large birefringence limit $|\delta_{HV}|\gg U_1,|\Delta_{H}|,\gamma,\gamma_B$, one has
\begin{eqnarray} \label{eq.C4}
g_{\sigma\sigma}^{(2)}=g_{\up\dn}^{(2)}&=&\frac{ \left(\gamma_B^2+4 \Delta_B^2\right) \left(\gamma ^2+4\Delta_H^2\right)}{ \gamma_B^2 \left( \gamma ^2+(2 \Delta_H+U_1/2)^2\right)+\left(g_{BL}^2-\Delta_{B} (4 \Delta_H+U_1)\right)^2+4 \gamma ^2 \Delta_{B}^2+2 \gamma  \gamma_B g_{BL}^2}.
\end{eqnarray}
Equations~\eqref{eq.C3} and \eqref{eq.C4} reduce to Eqs.~\eqref{eq:gresonanceNodhV} and \eqref{eq:gresonancedhV} given in the main text when the detunings are set to zero.

\twocolumngrid
\bibliography{biblio}
\end{document}